\pdfoutput=1

\documentclass[11pt,twoside,a4paper,cmspaper,final,collab]{cms-tdr}
\def\svnVersion{135ccdf-D}\def\svnDate{2020/08/26}\def\cmsCernNoTag{CERN-EP-2019-262}\def\cmsCernDate{\today}\def\cmsMessage{"Published in Physical Review D as \href{http://dx.doi.org/10.1103/PhysRevD.102.032007}{\doi{10.1103/PhysRevD.102.032007}.}"}

\begin{document}\cmsNoteHeader{SMP-19-004}

\hyphenation{had-ron-i-za-tion}
\hyphenation{cal-or-i-me-ter}
\hyphenation{de-vices}

\newcommand{\msv}{\ensuremath{M_\text{SV}}\xspace}
\newcommand{\zj}  {\ensuremath{\PZ\hspace{-.25em}+\hspace{-.15em}\text{jets}}\xspace}
\newcommand{\wj}  {\ensuremath{\PW\hspace{-.25em}+\hspace{-.15em}\text{jets}}\xspace}
\newcommand{\zc}  {\ensuremath{\PZ\hspace{-.25em}+\hspace{-.15em}\cPqc\text{~jet}}\xspace}
\newcommand{\wcs}  {\ensuremath{\PW\hspace{-.25em}+\hspace{-.15em}\cPqc\text{~jets}}\xspace}
\newcommand{\zqc} {\ensuremath{\PZ\hspace{-.25em}+\hspace{-.15em}\cPqc}\xspace}
\newcommand{\zb}  {\ensuremath{\PZ\hspace{-.25em}+\hspace{-.15em}\cPqb\text{~jet}}\xspace}
\newcommand{\zcs} {\ensuremath{\PZ\hspace{-.25em}+\hspace{-.15em}\cPqc\text{~jets}}\xspace}
\newcommand{\zbs} {\ensuremath{\PZ\hspace{-.25em}+\hspace{-.15em}\cPqb\text{~jets}}\xspace}
\newcommand{\zls} {\ensuremath{\PZ\hspace{-.25em}+\hspace{-.15em}\text{light~jets}}\xspace}
\newcommand{\zhfs}{\ensuremath{\PZ\hspace{-.25em}+\hspace{-.15em}\text{HF~jets}}\xspace}
\newcommand{\bj}{\cPqb~jet\xspace}
\newcommand{\cj}{\cPqc~jet\xspace}
\newcommand{\bjs}{\cPqb~jets\xspace}
\newcommand{\cjs}{\cPqc~jets\xspace}
\newcommand{\SFc}{\ensuremath{\mathrm{SF}_{\PQc}}\xspace}
\newcommand{\SFb}{\ensuremath{\mathrm{SF}_{\PQb}}\xspace}
\newcommand{\emu}{\ensuremath{\Pe\PGm}\xspace}
\newcommand{\mgfive}{\textsc{mg}5\_a\textsc{MC}\xspace}
\newcommand{\muR}{\ensuremath{\mu_\text{R}}\xspace}
\newcommand{\muF}{\ensuremath{\mu_\text{F}}\xspace}
\newcommand{\tuf}{\textsc{TUnfold}\xspace}
\ifthenelse{\boolean{cms@external}}{\providecommand{\cmsTable}[1]{#1}}{\providecommand{\cmsTable}[1]{\resizebox{\textwidth}{!}{#1}}}
\providecommand{\cmsTableRot}[1]{\resizebox{660pt}{!}{#1}}
\newlength\cmsTabSkip\setlength{\cmsTabSkip}{1ex}
\newcommand\T{\rule{0pt}{2.6ex}}
\newcommand\B{\rule[-1.2ex]{0pt}{0pt}}

\cmsNoteHeader{SMP-19-004}

\title{Measurement of the associated production of a \texorpdfstring{\PZ}{Z} boson with charm or bottom quark jets in proton-proton collisions at \texorpdfstring{$\sqrt{s}=13$\TeV}{sqrt(s)= 13 TeV}}

\date{\today}

\abstract{
Ratios of cross sections, $\sigma(\zcs)/\sigma(\zj)$, $\sigma(\zbs)/\sigma(\zj)$, and $\sigma(\zcs)/\sigma(\zbs)$ in the associated production of a \PZ boson with at least one charm or bottom quark jet are measured in proton-proton collisions at $\sqrt s=13$\TeV. The data sample, collected by the CMS experiment at the CERN LHC, corresponds to an integrated luminosity of 35.9\fbinv, with a fiducial volume of $\pt>30\GeV$ and $|\eta|<2.4$ for the jets, where \pt and $\eta$ represent transverse momentum and pseudorapidity, respectively. The \PZ boson candidates come from leptonic decays into electrons or muons with $\pt>25\GeV$ and $\abs{\eta}<2.4$, and the dilepton mass satisfies $71<m_{\PZ}<111\GeV$. The measured values are $\sigma(\zcs)/\sigma(\zj) = 0.102 \pm 0.002 \pm 0.009$, $\sigma(\zbs)/\sigma(\zj) = 0.0633 \pm 0.0004 \pm 0.0015$, and $\sigma(\zcs)/\sigma(\zbs) = 1.62 \pm 0.03 \pm 0.15$. Results on the inclusive and differential cross section ratios as functions of jet and \PZ boson transverse momentum are compared with predictions from leading and next-to-leading order perturbative quantum chromodynamics calculations. These are the first measurements of the cross section ratios at 13\TeV.
}

\hypersetup{%
pdfauthor={CMS Collaboration},%
pdftitle={Measurement of the associated production of a Z boson with charm or bottom
quark jets in proton-proton collisions at sqrt(s) = 13 TeV},%
pdfsubject={CMS},%
pdfkeywords={CMS, physics, jets}}

\maketitle

\section{Introduction}\label{intro}

Studies of \PZ boson production in association with heavy-flavor (HF) jets from the hadronization of heavy (\cPqc or \cPqb) quarks provide important tests of perturbative quantum chromodynamics (pQCD) calculations. A good description of these processes is also important since they form a major background for a variety of physics processes including Higgs boson production in association with a \PZ boson, $\PZ\PH$ ($\PH\to\cPqc\cPaqc$ or $\PH\to\cPqb\cPaqb$), and searches for new physics signatures in final states with leptons and HF jets. Two different approaches are currently available for calculating the \zhfs production: the five-flavor scheme (5FS)~\cite{Campbell:2003dd} and the four-flavor scheme (4FS)~\cite{Cordero:2009kv}. Both approaches yield consistent results within theoretical uncertainties~\cite{Maltoni}.

Several \zhfs measurements have been performed by the CDF and D0 Collaborations at the FNAL Tevatron~\cite{CDF,D02,D03} and by the ATLAS, CMS, and LHCb Collaborations at the CERN LHC~\cite{ATLAS,CMS1,LHCb1,LHCb2}.
The D0 Collaboration reported on the first $\sigma(\zcs)/\sigma(\zbs)$ cross section ratio measurement \cite{D02} and observed a significantly higher value compared to next-to-leading order (NLO) pQCD calculations. A measurement of the $\sigma(\zcs)/\sigma(\zbs)$ cross section ratio in 8\TeV proton-proton ($\Pp\Pp$) collisions at the LHC has been recently reported by the CMS Collaboration~\cite{ZCciemat} and is in agreement with predictions from leading order (LO) and NLO calculations obtained with the \MADGRAPH~\cite{Alwall:2011uj} and \MGvATNLO~\cite{mg5} programs, respectively.

The current paper reports on simultaneous measurements of the \cPqc and \cPqb quark jet contents in a sample containing a \PZ boson (in the following, \PZ is used as a shorthand for \PZ/$\gamma^*$) produced in association with at least one jet. These measurements provide the first results for proton-proton collisions at $\sqrt s=13$\TeV. The experimental precision is improved with respect to previous LHC results because of the increased size of the data sample and advanced heavy-flavor tagging techniques. The \PZ bosons are identified through reconstructed dielectrons or dimuons, where the individual leptons are subject to requirements on transverse momentum ($\pt > 25\GeV$) and pseudorapidity ($\abs{\eta} < 2.4$). The dilepton invariant mass must be within a \PZ boson window of 71--111\GeV, and jets are required to have $\pt > 30\GeV$ and $\abs{\eta} < 2.4$.

The following cross section ratios are measured: $\sigma(\zcs)/\sigma(\zj)$, $\sigma(\zbs)/\sigma(\zj)$, and $\sigma(\zcs)/\sigma(\zbs)$. These cross section ratios are measured inclusively and differentially as functions of the transverse momentum of the jet and the \PZ boson, and are unfolded to the particle level taking into account detector effects.
The measurements of the cross section ratios benefit from cancellations of several systematic uncertainties related to the jet, lepton, and luminosity measurements. A number of theory-related uncertainties are reduced as well, including those linked to the details of parton showering and hadronization. Therefore, by measuring cross section ratios one can more precisely compare data with theoretical calculations.

The paper is organized as follows. The CMS experiment and data together with simulated samples used in the analysis are described in Sections~\ref{sec:dec} and~\ref{sec:sample}. Details of the measurements are described in Sections~\ref{sec:sel},~\ref{sec:measurement}, and~\ref{sec:unfolding}, while Sections~\ref{sec:syst} and~\ref{sec:result} present the systematic uncertainties and the measurement results, respectively, followed by a summary in Section~\ref{sec:sum}.

\section{The CMS detector}\label{sec:dec}

The central feature of the CMS apparatus is a superconducting solenoid of 6\unit{m} internal diameter, providing a magnetic field of 3.8\unit{T}. Within the solenoid volume are a silicon pixel and strip tracker, covering a pseudorapidity region of $\abs{\eta}<2.5$, a lead tungstate crystal electromagnetic calorimeter (ECAL), and a brass and scintillator hadron calorimeter, each composed of a barrel and two endcap sections. Forward calorimeters, made of steel and quartz fibers, extend the pseudorapidity coverage provided by the barrel and endcap detectors to $\abs{\eta}<5$. Muons are detected in gas-ionization chambers embedded in the steel flux-return yoke outside the solenoid and covering $\abs{\eta}<2.4$.

Events of interest are selected using a two-tiered trigger system~\cite{Khachatryan:2016bia}. The first level, composed of custom hardware processors, uses information from the calorimeters and muon detectors to select events at a rate of around 100\unit{kHz} within a time interval of less than 4\mus. The second level, known as the high-level trigger, consists of a farm of processors running a version of the full event reconstruction software optimized for fast processing, and reduces the event rate to around 1\unit{kHz} before data storage.

A more detailed description of the CMS detector, together with a definition of the coordinate system used and the relevant kinematic variables, can be found in Ref.~\cite{Chatrchyan:2008zzk}.

\section{Data and simulated samples}\label{sec:sample}

The cross section ratio measurements are based on proton-proton ($\Pp\Pp$) collision data at $\sqrt{s}=13\TeV$ collected with the CMS detector in 2016 and corresponding to an integrated luminosity of 35.9\fbinv~\cite{lumi}. Recorded events have an average 23 additional $\Pp\Pp$ interactions per bunch crossing (pileup) together with the hard process.

Various Monte Carlo (MC) event generators are used to simulate the \zj signal and background processes. The full detector simulation is based on the \GEANTfour package~\citep{geant4}. The simulation includes the pileup effects from the same or nearby bunch crossings by overlapping the hard process of interest with the pileup events. The simulated events are reconstructed with the same algorithms as used for the data.

The \zj events are generated by \MGvATNLO v2.2.2~\cite{mg5} (using 5FS; denoted as \mgfive in the following) at NLO in pQCD with up to two additional partons at the matrix element level, generated for each parton multiplicity and then merged. The \mgfive matrix element generator is interfaced with \PYTHIA v8.212~\cite{Sjostrand:2014zea}, which simulates the parton shower and hadronization processes, through the FxFx merging scheme~\cite{Frederix:2012ps} at a matching scale of 19\GeV. The predicted numbers of events from \zj production are estimated using a cross section at next-to-next-to-leading order (NNLO) accuracy obtained from \FEWZ v3.1~\cite{Melnikov:2006kv}.

The background events originate from top quark and diboson processes. Top quark-antiquark (\ttbar) production, which forms the dominant background, is generated at NLO by \POWHEG v2.0~\cite{powheg1,powheg2,powheg3,powheg4} and normalized to a cross section calculated by using~\textsc{top++}~v2.0~\cite{top++} at NNLO accuracy including soft-gluon resummation. The diboson ($\PW\PW$, $\PW\PZ$, $\PZ\PZ$) backgrounds are generated by \PYTHIA while \POWHEG and NLO \mgfive are used to simulate the single top quark processes ($s$-channel, $t$-channel, and $\cPqt\PW$). The \POWHEG generator is also interfaced with \PYTHIA for parton showering and hadronization. The diboson and single top quark predictions are normalized to NNLO~\cite{Grazzini:2017mhc,Kidonakis:2013zqa} cross sections.

The NNPDF 3.0 NLO and LO parton distribution functions (PDF)~\cite{Ball:2014uwa} are used for the NLO and LO calculations, respectively. \PYTHIA uses the NNPDF 2.3 LO PDF set and the CUETP8M1~\cite{Khachatryan:2015pea} or CUETP8M2T4~\cite{CMS-PAS-TOP-16-021} (\ttbar sample) for the underlying event tune.

\section{Object reconstruction and event selection}\label{sec:sel}

The particle-flow (PF) algorithm~\cite{CMS-PRF-14-001} reconstructs and identifies each individual particle in an event, with an optimized combination of information from the various elements of the CMS detector. The neutral particle energy deposits are determined in the calorimeters, whereas charged tracks are measured in the central tracking and muon systems.

The candidate vertex with the largest value of summed physics-object $\pt^2$ is taken to be the primary $\Pp\Pp$ interaction vertex. The physics objects are the jets, clustered using the jet finding algorithm~\cite{Cacciari:2008gp,Cacciari:2011ma} with the tracks assigned to candidate vertices as inputs, and leptons. More details are given in Ref.~\cite{CMS-TDR-15-02}.%and the associated missing transverse momentum, taken as the negative vector sum of the \pt of those jets

Electrons are reconstructed using momentum measurements in the tracker combined with the energy deposits in the ECAL~\cite{Khachatryan:2015hwa}. The identification requirements are based on the ECAL shower shape, matching between the electron track and the energy clusters in the ECAL, and observables characterizing the bremsstrahlung along the electron trajectory. Electrons are required to originate from the primary vertex. The electron momentum is estimated by combining the energy measurement in the ECAL with the momentum measurement in the tracker. The momentum resolution for electrons with $\pt \approx 45\GeV$ from $\PZ \to \Pep \Pem$ decays ranges from 1.7 to 4.5\%. The resolution tends to be better in the barrel region than in the endcaps, and it also depends on the bremsstrahlung energy emitted by the electron as it traverses the material in front of the ECAL. The dielectron mass resolution for $\PZ \to \Pep \Pem$ decays when both electrons are in the ECAL barrel is 1.9\%, and is 2.9\% when both electrons are in the endcaps~\cite{Khachatryan:2015hwa}.

Muon candidates are built by combining signals from the tracker and the muon subsystems. The identification criteria are based on the number of measurements in the detectors, the fit quality of the track, and requirements on its association with the primary vertex. Matching muons to tracks measured in the tracker results in a relative transverse momentum resolution, for muons with \pt up to 100\GeV, of 1\% in the barrel and 3\% in the endcaps~\cite{Sirunyan:2018}.

To reduce the misidentification rate, electrons and muons are required to be isolated.  Activity near an electron (muon) is quantified as the sum of transverse momenta of PF candidates within the isolation cone radius of $\Delta R=\sqrt{\smash[b]{(\Delta\eta)^2+(\Delta\phi)^2}}=0.3$~(0.4) around the electron (muon) track, where $\phi$ is the azimuthal angle. After compensating for the energy contribution from pileup in the isolation cone, the resultant sum is required to be less than 25\% of the lepton transverse momentum. The lepton isolation, along with other requirements to select \zj events, strongly suppresses background events with misidentified dileptons such as \wj and QCD multijets.

Based on the PF candidates, jets are reconstructed using the anti-\kt algorithm with a distance parameter of 0.4. Jet momentum is determined as the vector sum of all particle momenta in the jet; based on simulation this is, on average, within 5 to 10\% of the true jet momentum over the entire \pt spectrum and detector acceptance. Pileup interactions can result in more tracks and calorimetric energy depositions, increasing the apparent jet momentum. To mitigate this effect, tracks originating from pileup vertices are discarded and an offset correction is applied to account for remaining contributions~\cite{jetPU1,jetPU2}. Jet energy corrections are derived from simulation studies so that the average measured response of jets becomes identical to that of particle-level jets. In situ measurements of the momentum balance in dijet, photon+jet, $\PZ$+jet, and multijet events are used to determine any residual differences between the jet energy scale (JES) in data and in simulation, and appropriate corrections are applied~\cite{Khachatryan:2016kdb}.  The jet energy resolution (JER) typically amounts to 16\% at 30\GeV and 8\% at 100\GeV. Additional selection criteria are applied to remove jets potentially dominated by instrumental effects or reconstruction failures~\cite{CMS-PAS-JME-16-003}.

For this analysis, the selection for \zj events starts with the trigger requirements based on two electron (muon) objects identified by the trigger system that pass \pt thresholds of 23 and 13\GeV (17 and 8\GeV). The \zj events are further selected by requiring two reconstructed electrons or muons with $\pt > 25\GeV$ and within $\abs{\eta}<2.4$. The \pt requirement is chosen to obtain high trigger efficiency for selecting the signal events. Events containing two selected electrons (muons) are categorized in the electron (muon) channel. The lepton candidates are subject to requirements on their transverse impact parameter, $\abs{d_{xy}} < 0.05\unit{cm}$, and their longitudinal impact parameter, $\abs{d_z} < 0.2\unit{cm}$, both with respect to the primary vertex. The \PZ boson candidate is reconstructed from a pair of oppositely charged same-flavor leptons with invariant mass between 71 and 111\GeV. An event must contain at least one associated jet with $\pt > 30\GeV$ and $\abs{\eta}<2.4$.

Missing transverse momentum is used in this analysis to reduce background contributions from \ttbar and single top quark production processes. In contrast to the \zj, these processes have a significant amount of missing energy because of undetected neutrinos in top quark decays. The missing transverse momentum vector, \ptvecmiss, is computed as the negative vector sum of the transverse momenta of all PF candidates in an event~\cite{CMS-PAS-JME-17-001_paper} and is further modified to account for corrections to the energy scale of the reconstructed jets. Its magnitude, \ptmiss, is required to be less than 40\GeV, which results in a signal efficiency of ${\approx}80\%$ with the \ttbar rejection factor of ${\approx}4.8$.

A \zj sample with enriched \cPqc and \cPqb quark jet content is selected by applying an HF tagging requirement to jets in the \zj sample described above. The discrimination of HF jets from light-flavor quark and gluon jets, referred to as light jets in the following, is achieved by constructing a discriminator variable from tracks and secondary vertex (SV) characteristics. Artificial neural network algorithms are used to combine specific properties of the HF quarks, long lifetime and substantial mass, to build the discriminator. The algorithm used in the analysis, the combined secondary vertex (Version 2), is described in Ref.~\cite{csv}. Some of the important input variables are the number of secondary vertices and the number of tracks associated with each of them, the mass and 2D decay distance significance of the SV with the smallest decay distance uncertainty, and the signed 3D impact parameter significance of the tracks. Here the significance is defined as the ratio between a measured quantity and its uncertainty. Although the combined secondary vertex (Version 2) is trained to distinguish \bjs, it does occasionally tag a \cj. Therefore, at a proper operating point, the algorithm can retain a sufficient amount of \cjs while heavily suppressing light jets. The analysis uses a ``medium" operating point, which corresponds to approximately 10 (60)\% tagging efficiencies for \cPqc (\cPqb) quark jets and a misidentification probability of 1\% for a light jet. The \zhfs sample must contain at least one tagged jet. The tagging efficiencies are determined using MC samples and corrected for the difference between data and simulation. The corresponding correction factors are derived from the data versus simulation efficiency comparisons in dedicated control samples containing \ttbar and multijet events~\cite{csv}.

In simulation, the classification of reconstructed \zj events into \zcs, \zbs, and \zls categories is based on the flavors of reconstructed jets with $\pt > 30\GeV$ and $\abs{\eta}<2.4$. They are classified as \cPqc or \cPqb jets if they are matched to MC generated \cPqc or \cPqb hadrons~\cite{csv}. In the case where both \cPqc and \cPqb hadrons are matched, the jet is considered a \bj. Based on reconstructed jets with defined flavors, events are classified as \zbs if they contain at least one \bj. Of the remaining events, those that contain at least one \cPqc hadron are considered as \zcs and those that contain neither \cPqc nor \cPqb hadrons are classified as \zls.

Table~\ref{tab:event_counts} lists the number of events estimated in simulation and found in data that satisfy the \zj and \zhfs selection criteria for both the electron and muon channels. The background, mostly from top quark and diboson processes, is approximately 5\% in the \zhfs sample, and the diboson background is dominated by the $\PW\PZ$ events. 

\begin{table*}[hbt]
  \topcaption{Numbers of events that satisfy the \zj and \zhfs selection criteria in the electron and muon channels. The uncertainties are statistical only.}
  \label{tab:event_counts}
  \centering
  \begin{scotch}{lcccc}
   & \multicolumn{2}{c}{\zj sample} & \multicolumn{2}{c}{\zhfs sample}\\
   &                     Electron &          Muon &          Electron            & Muon\\
  \hline
  \zcs & 171\,970 $\pm$ 530   & 287\,090 $\pm$ 720  & 18\,870 $\pm$ 170 & \xspace\xspace 32\,310 $\pm$ 230\\
  \zbs & \xspace\xspace 95\,910 $\pm$ 410 & 159\,500 $\pm$ 560   & 60\,100 $\pm$ 310 & 100\,630 $\pm$ 420\\
  \zls &              1\,531\,900 $\pm$ 1\,600 & 2\,612\,100 $\pm$ 2\,200 & \xspace\xspace 6\,170 $\pm$ 100 & \xspace\xspace 10\,810 $\pm$ 140\\
  \ttbar & \xspace\xspace 5\,850 $\pm$ 50      & \xspace \xspace 9\,440 $\pm$ 60 & 3\,850 $\pm$ 40 & \xspace\xspace 6\,180 $\pm$ 50\\
  Diboson & 10\,040 $\pm$ 60      & 16\,310 $\pm$ 80     & \xspace\xspace\xspace 780 $\pm$ 20 & \xspace\xspace 1\,320 $\pm$ 20\\
  Single \cPqt & \xspace\xspace\xspace\xspace\xspace 580 $\pm$ 10       & \xspace\xspace\xspace\xspace\xspace 950 $\pm$ 10 & \xspace 303 $\pm$ 7 & \xspace\xspace\xspace\xspace 500 $\pm$ 10\\
  [\cmsTabSkip]
  Total, simulation &          1\,816\,200 $\pm$ 1\,700 & 3\,085\,400 $\pm$ 2\,400 & 90\,070 $\pm$ 370 & 151\,740 $\pm$ 510\\
  Data &              1\,759\,047            & 2\,959\,629            & 79\,015 &   130\,775 \\
  Data/simulation & \xspace\xspace\xspace\xspace\xspace\xspace\xspace 0.969 $\pm$ 0.001  & \xspace\xspace\xspace\xspace\xspace\xspace\xspace 0.959 $\pm$ 0.001  & \xspace\xspace\xspace\xspace\xspace 0.877 $\pm$ 0.005 & \xspace\xspace\xspace\xspace\xspace\xspace\xspace 0.862 $\pm$ 0.004\\
  \end{scotch}
\end{table*}

\section{Cross section ratio measurements}\label{sec:measurement}
\subsection{Analysis strategy}\label{subsec:strategy}

The goal of the analysis is to precisely measure the fraction of jets with heavy flavors in \zj events. For this purpose, the SV invariant mass, \msv, of the tagged jet with highest \pt in the \zhfs events is used. The SV is reconstructed using an adaptive vertex reconstruction algorithm~\cite{Waltenberger:1166320} from selected tracks within a cone of $\Delta R < 0.3$ around the jet axis. The distance between the track and the jet axis measured at their point of closest approach must be less than 0.2\cm. Details of track selections and SV reconstructions can be found in Refs.~\cite{BTV12001,csv}. The \msv is calculated using the momenta of charged-particle tracks associated with the SV. The corresponding particles are assumed to have the pion mass for the purpose of calculating the SV mass. The \msv distributions possess specific features depending on the jet flavor, and can be used as templates in a fit to the \msv distribution in data to extract the fractions of \cPqc and \cPqb jets, as discussed in Section~\ref{sec:sf}.

The template fit is performed in the \zj sample enriched with HF jets, \ie, in the \zhfs sample, and the observed number of \zcs ($N_{\PQc}$) and \zbs ($N_{\PQb}$) events are derived. They are corrected for the efficiencies of tagging events, $\epsilon^{\PQc}_{\text{tag}}$ and $\epsilon^{\PQb}_{\text{tag}}$ for $N_{\PQc}$ and $N_{\PQb}$, respectively, to obtain the numbers of \zcs and \zbs events in the \zj sample. The cross section ratios are then calculated as
\begin{equation}\label{eq:rcj}
  \text{R}(\cPqc/\text{j}) = \frac{\sigma(\zcs)}{\sigma(\zj)} = \frac{N_{\PQc}}{N_{\text{jet}}\epsilon^{\PQc}_{\text{tag}}},
\end{equation}
\begin{equation}\label{eq:rbj}
  \text{R}(\cPqb/\text{j}) = \frac{\sigma(\zbs)}{\sigma(\zj)} = \frac{N_{\PQb}}{N_{\text{jet}}\epsilon^{\PQb}_{\text{tag}}},
\end{equation}
\begin{equation}\label{eq:rcb}
  \text{R}(\cPqc/\cPqb) = \frac{\sigma(\zcs)}{\sigma(\zbs)} = \frac{N_{\PQc}\epsilon^{\PQb}_{\text{tag}}}{N_{\PQb}\epsilon^{\PQc}_{\text{tag}}},
\end{equation}
where $N_{\text{jet}}$ is the number of selected \zj events remaining after subtracting background contributions (\ttbar, diboson, and single top) from data. These backgrounds are estimated using simulation. In the above formulas, the integrated luminosity as well as the efficiencies that are related to lepton and \ptmiss requirements in the \zj event selection cancel.

For the differential measurements, the same procedure described here is applied in each jet or \PZ boson \pt interval. Dedicated \msv templates are derived for each interval to take into account the dependence of the \msv shape on jet kinematic variables. Finally, the cross section ratios are unfolded for various experimental effects, most notably the detector resolution and efficiencies.

\subsection{\texorpdfstring{\PZ{}+HF jets}{Z+HF Jets} event tagging efficiency}\label{subsec:eff}

The efficiencies of tagging \zhfs events, $\epsilon^{\PQc}_{\text{tag}}$ and $\epsilon^{\PQb}_{\text{tag}}$, are calculated as the ratio between numbers of selected \zcs and \zbs events, respectively, in the \zhfs and the \zj samples. They are estimated using simulations, which are corrected with data. In the jet \pt range between 30 and 200\GeV the efficiencies vary only slightly and range from 8.3 to 11.3\% for \zcs and from 45.9 to 60.7\% for \zbs events. The \zls mistagging rate increases from 0.3 to 1.0\% in the same \pt range.

\subsection{Estimation of the event yields}\label{sec:sf}
A binned maximum likelihood template fit, based on \msv distributions of the leading \pt HF-tagged jets, is used to obtain the numbers of \zcs and \zbs events in the \zhfs sample. The parameters of interest are the scale factors, \SFc and \SFb, that adjust the MC rates to fit the data, while their uncertainties are treated as nuisance parameters. The \msv distributions of the simulated \zcs, \zbs, and \zls categories are normalized to the integrated luminosity of the data sample using an NNLO cross section for the total \zj rate. The top quark and diboson backgrounds, which contribute about 5\% of the events in the \zhfs sample, are estimated from simulation. The predicted yields of all these processes are shown in Table~\ref{tab:event_counts}.

For each \msv bin a Poisson distribution is constructed from the number of observed events, with its mean taken from MC predictions of signal (\zcs and \zbs) and background (\zls, top quark, and diboson) yields. The likelihood is the product of the Poisson distributions and Gaussian (or log-normal) distributions, where the latter are used to constrain the nuisance parameters. The choice of Gaussian or log-normal constraints depends on whether the corresponding systematic uncertainty affects the shape or normalization of the templates, respectively. To obtain a combined result, the electron and muon channel data are fitted simultaneously using a common set of scale factors. After the fit, the numbers of \zcs and \zbs events, $N_{\PQc}$ and $N_{\PQb}$, are obtained from the MC predictions scaled by the \SFc and \SFb factors.

The \msv template of \cjs in the \zcs events is obtained from simulation. The \cj \msv shape is validated with a \ttbar-enriched data sample where only one of the \PW bosons decays to leptons. The other \PW boson decays hadronically with a branching fraction of 33\% for a charm quark in the final state. The event selection requires a well-identified and isolated muon having $\pt > 25\GeV$ and $\abs{\eta} < 2.4$ together with at least four jets, each with $\pt >30\GeV$ and $\abs{\eta} < 2.4$. The \cPqc and \cPqb jet identification is performed with the following procedure. To reduce the combinatorics the best pair (triplet) of jets is chosen by minimizing the reconstructed and nominal mass of the \PW boson (top quark). From this optimization, the \cPqc and \bj candidates from top quark and \PW boson decays are identified. The event is kept if these candidates pass the jet HF tagging requirement described in Section~\ref{sec:sel}. In the resulting sample of the \cj candidates about half have correct flavor assignment whereas the other half constitute mostly \bjs that are misidentified as \cjs. The \cj \msv distribution in data is found by subtracting the backgrounds containing \bjs and light jets in the sample.

\begin{figure*}[tb]
  \centering
  \includegraphics[width=0.4\textwidth]{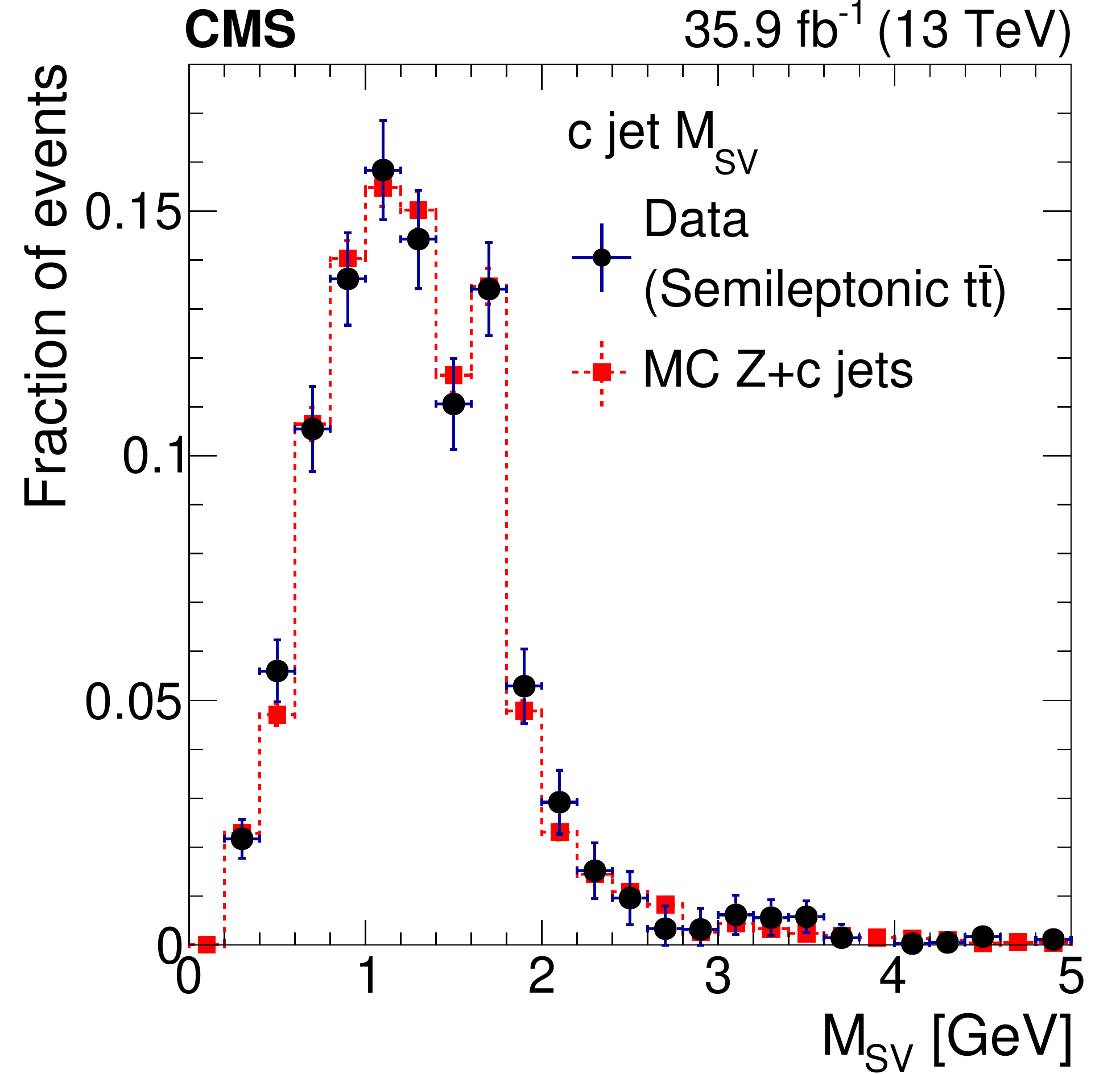}
  \includegraphics[width=0.4\textwidth]{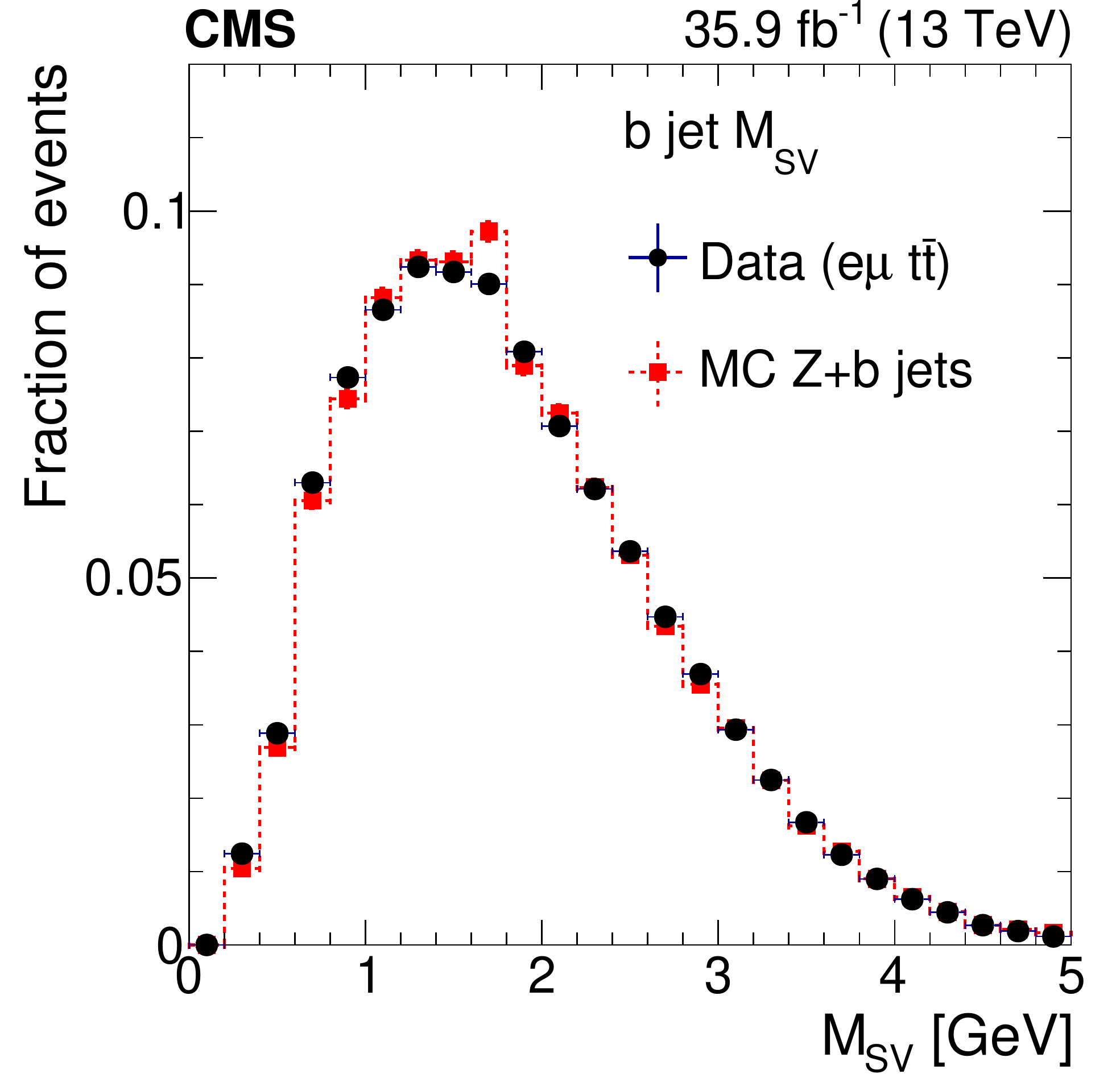}
  \caption{Comparisons of \cj (left) and \bj (right) \msv distributions for data and simulation. A shape correction is applied to the data-driven distribution in the right panel to account for the difference between the jets in \ttbar and \zbs processes.}
  \label{fig:temp_val}
\end{figure*}

The \cj \msv template from simulation is compared with that observed in the validation sample and agreement is found within the statistical uncertainties as shown in Fig.~\ref{fig:temp_val} (left). The pronounced enhancement seen in the \cjs \msv distribution near 1.8\GeV is due to charm meson decays.

The \msv template for \bjs is derived from a high-purity data sample of \ttbar events decaying to final states of \emu $+\ge$ 2 jets with at least one b-tagged jet. Leptons must pass similar requirements as those used in the selection of \zj events except for a tighter isolation criterion (other activity with less than 15\% of the lepton transverse momentum, instead of 25\%) to strongly suppress multijet and \wj backgrounds. The \msv shape depends on the kinematic distributions of the corresponding jets, therefore the \bj \msv templates obtained with the \ttbar data are corrected to account for the difference between the \bj \pt spectra in \ttbar and \zbs events. This correction is derived from simulation by comparing the \bj \msv shapes in those two samples of events. It is parameterized as a second-order polynomial function of \msv and varies between 3 and 20\% across jet \pt ranges. A comparison between the simulated and data-driven \bj \msv distributions are presented in Fig.~\ref{fig:temp_val} (right). This correction procedure is applied in both the inclusive and differential measurements.

The \msv modeling of light jets in simulation is checked in the validation sample containing \wj events selected by requiring a well-identified and isolated muon together with at least one jet. Discrimination criteria of \cj versus light jets are applied, resulting in a sample with light jet purity of ${\approx}40\%$. The light jet \msv templates in data are derived from the validation sample after subtracting nonlight jet components, which mainly consist of the \wcs events. Good agreement between the data-driven template shape and the simulation is observed.

\begin{figure*}[htb]
\centering
\includegraphics[width=0.45\textwidth]{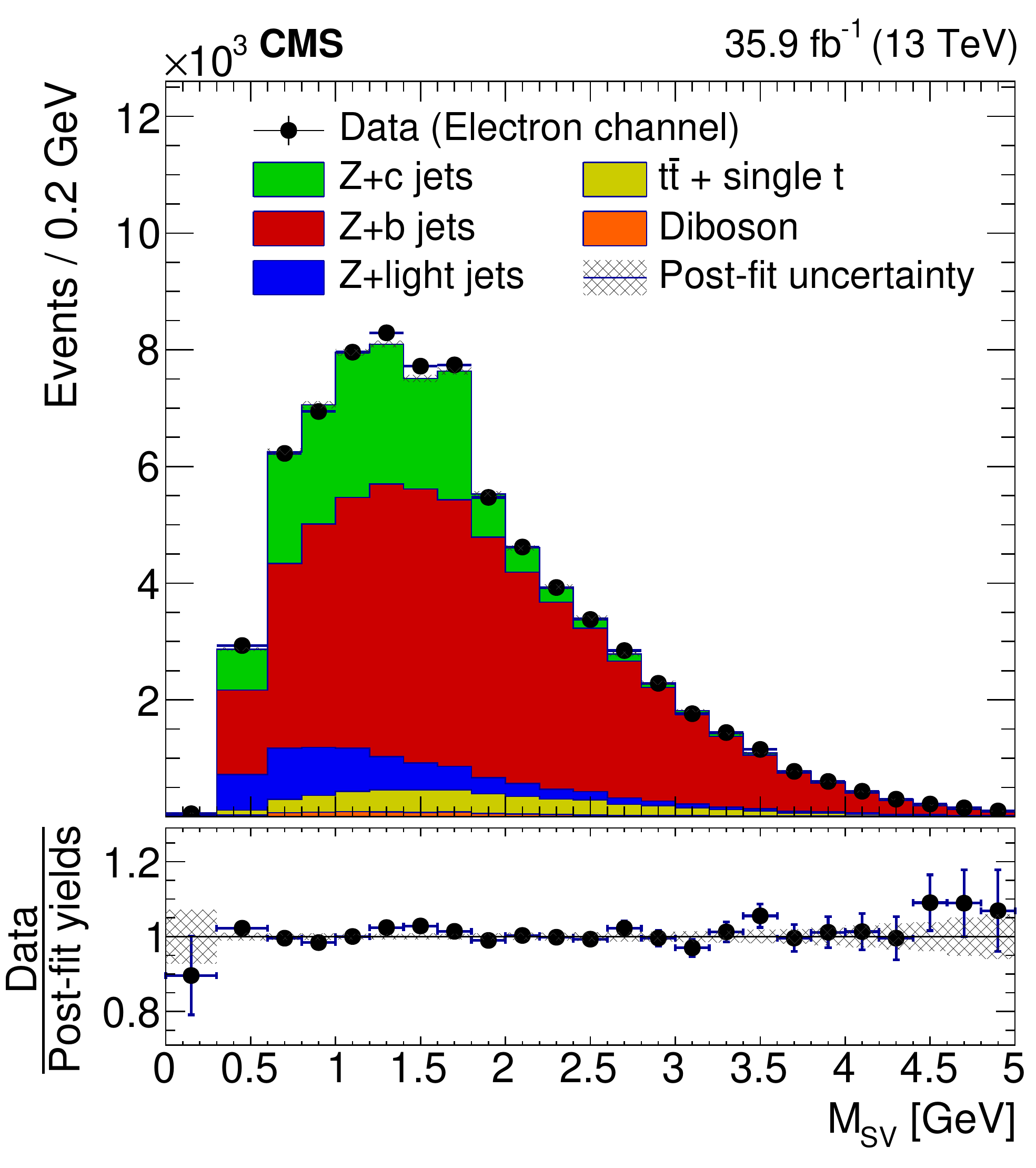}
\includegraphics[width=0.45\textwidth]{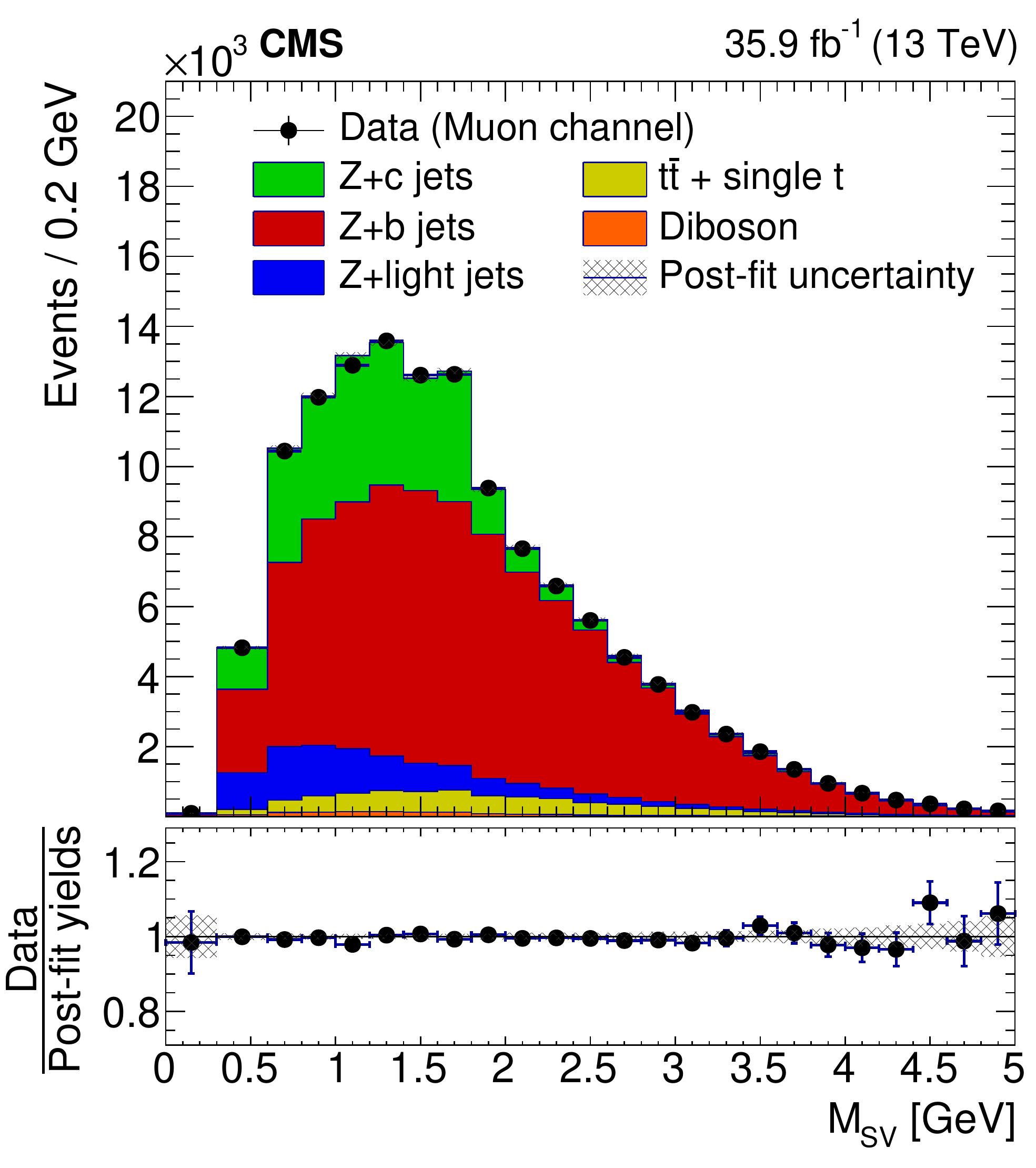}
\caption{Secondary vertex invariant mass distributions for the electron (left) and muon (right) channels derived from fits using the inclusive \zhfs data sample. The post-fit uncertainty bands indicate the total uncertainties, added in quadrature, of the best-fit values of signal and background process rates.}
\label{fig:fig_postfit_1}
\end{figure*}

The scale factors obtained from the combined fit in the inclusive \zhfs data sample are $\text{SF}_{\text{c}}=0.849 \pm 0.013 \text{ (stat)} \pm 0.064 \text{ (syst)}$ and $\text{SF}_{\text{b}}=0.873 \pm 0.005 \text{ (stat)} \pm 0.013 \text{ (syst)}$. Tables~\ref{tab:sf_fit_1} and~\ref{tab:sf_fit_2} list the scale factors estimated in the jet and \PZ \pt bins. Details on the evaluations of systematic uncertainties in the scale factors are discussed in Section~\ref{sec:syst}. The two channels pass a $\chi^2$ based compatibility check except for the \SFc fluctuation in one jet \pt bin of 50--110\GeV with a $p$-value of ${\approx}0.3\%$. The post-fit \msv distributions are shown in Fig.~\ref{fig:fig_postfit_1} for the measurements using the inclusive \zhfs samples. Examples of the post-fit \msv distributions in the muon channel for exclusive jet \pt bins are given in Appendix~\ref{sec:msv_dist}, Fig.~\ref{fig:fig_postfit_2}.

\begin{table*}[ht]
\topcaption{The \SFc and \SFb scale factor fit results for electron, muon, and combined channels in jet \pt bins. The first and second uncertainty values correspond to the statistical and systematic contributions, respectively. The fractions of the observed number of \zcs and \zbs in the total number of \zj events selected in the \zhfs sample are shown in the parentheses and are determined by applying the scale factors to the corresponding MC events.}
\label{tab:sf_fit_1}
\centering
\cmsTable{
\begin{scotch}{ccccccc}
\multirow{3}{*}{\begin{tabular}{@{}c@{}} Jet \pt bins \\ (\GeVns) \end{tabular} } &\multicolumn{3}{c}{\SFc} & \multicolumn{3}{c}{\SFb}\\
&\multicolumn{3}{c}{(\zc fraction)} & \multicolumn{3}{c}{(\zb fraction)}\\
& Electron & Muon & Combined & Electron & Muon & Combined\\
\hline
30--35&0.91$\pm$0.05$\pm$0.07 & 0.88$\pm$0.03$\pm$0.07 & 0.89$\pm$0.03$\pm$0.06 & 0.83$\pm$0.02$\pm$0.03 & 0.83$\pm$0.01$\pm$0.03 & 0.83$\pm$0.01$\pm$0.03\\
 & (25.4$\pm$1.3$\pm$2.1)\% & (25.8$\pm$1.0$\pm$2.1)\% & (25.6$\pm$0.8$\pm$1.7)\% & (64.8$\pm$1.4$\pm$2.4)\% & (64.2$\pm$1.1$\pm$2.4)\% & (64.5$\pm$0.9$\pm$1.9)\%\\
35--40&0.78$\pm$0.05$\pm$0.08 & 0.79$\pm$0.04$\pm$0.07 & 0.79$\pm$0.03$\pm$0.06 & 0.91$\pm$0.02$\pm$0.03 & 0.87$\pm$0.01$\pm$0.03 & 0.89$\pm$0.01$\pm$0.03\\
 & (21.1$\pm$1.4$\pm$2.0)\% & (22.3$\pm$1.1$\pm$2.0)\% & (21.7$\pm$0.9$\pm$1.7)\% & (69.7$\pm$1.5$\pm$2.6)\% & (68.7$\pm$1.2$\pm$2.4)\% & (69.2$\pm$0.9$\pm$2.1)\%\\
40--50&0.66$\pm$0.04$\pm$0.06 & 0.67$\pm$0.03$\pm$0.07 & 0.67$\pm$0.03$\pm$0.06 & 0.83$\pm$0.01$\pm$0.02 & 0.83$\pm$0.01$\pm$0.02 & 0.83$\pm$0.01$\pm$0.02\\
 & (18.2$\pm$1.2$\pm$1.8)\% & (18.4$\pm$0.9$\pm$1.8)\% & (18.3$\pm$0.7$\pm$1.7)\% & (73.9$\pm$1.3$\pm$1.9)\% & (73.1$\pm$1.1$\pm$1.8)\% & (73.4$\pm$0.8$\pm$1.6)\%\\
50--110&0.89$\pm$0.04$\pm$0.06 & 0.73$\pm$0.03$\pm$0.05 & 0.79$\pm$0.02$\pm$0.05 & 0.89$\pm$0.01$\pm$0.02 & 0.92$\pm$0.01$\pm$0.03 & 0.91$\pm$0.01$\pm$0.03\\
 & (20.3$\pm$0.8$\pm$1.3)\% & (17.5$\pm$0.6$\pm$1.2)\% & (18.6$\pm$0.5$\pm$1.1)\% & (72.9$\pm$1.0$\pm$1.8)\% & (75.1$\pm$0.7$\pm$2.5)\% & (74.1$\pm$0.6$\pm$2.3)\%\\
110--200&0.70$\pm$0.09$\pm$0.06 & 0.85$\pm$0.07$\pm$0.07 & 0.79$\pm$0.06$\pm$0.05 & 0.92$\pm$0.04$\pm$0.04 & 0.81$\pm$0.03$\pm$0.04 & 0.84$\pm$0.02$\pm$0.04\\
 & (18.0$\pm$2.3$\pm$1.6)\% & (21.4$\pm$1.8$\pm$1.7)\% & (20.2$\pm$1.4$\pm$1.4)\% & (71.5$\pm$2.7$\pm$3.1)\% & (67.1$\pm$2.2$\pm$3.2)\% & (68.7$\pm$1.7$\pm$3.0)\%\\
\end{scotch}
}
\end{table*}
\begin{table*}[ht]
\topcaption{The \SFc and \SFb scale factor fit results for electron, muon, and combined channels in \PZ \pt bins. The first and second uncertainty values correspond to the statistical and systematic contributions, respectively. The fraction of the observed number of \zcs and \zbs in the total number of \zj events selected in the \zhfs sample are shown in the parentheses and are derived by applying the scale factors to the corresponding MC events.}
\label{tab:sf_fit_2}
\centering
\cmsTable{
\begin{scotch}{ccccccc}
\multirow{3}{*}{\begin{tabular}{@{}c@{}} \PZ \pt bins \\ (\GeVns) \end{tabular}} &\multicolumn{3}{c}{\SFc} & \multicolumn{3}{c}{\SFb}\\
&\multicolumn{3}{c}{(\zc fraction)} & \multicolumn{3}{c}{(\zb fraction)}\\
& Electron & Muon & Combined & Electron & Muon & Combined\\
\hline
0--30&0.83$\pm$0.04$\pm$0.12 & 0.78$\pm$0.03$\pm$0.10 & 0.80$\pm$0.03$\pm$0.11 & 0.93$\pm$0.02$\pm$0.02 & 0.91$\pm$0.01$\pm$0.02 & 0.92$\pm$0.01$\pm$0.02\\
 & (22.9$\pm$1.1$\pm$3.3)\% & (21.9$\pm$0.9$\pm$2.9)\% & (22.4$\pm$0.7$\pm$3.1)\% & (67.9$\pm$1.2$\pm$1.7)\% & (68.1$\pm$1.0$\pm$1.7)\% & (67.9$\pm$0.7$\pm$1.6)\%\\
30--50&0.79$\pm$0.04$\pm$0.07 & 0.72$\pm$0.03$\pm$0.06 & 0.75$\pm$0.02$\pm$0.06 & 0.84$\pm$0.01$\pm$0.02 & 0.84$\pm$0.01$\pm$0.02 & 0.84$\pm$0.01$\pm$0.02\\
 & (21.6$\pm$1.0$\pm$1.9)\% & (20.5$\pm$0.8$\pm$1.8)\% & (20.9$\pm$0.6$\pm$1.8)\% & (71.0$\pm$1.1$\pm$1.4)\% & (71.6$\pm$0.9$\pm$1.4)\% & (71.4$\pm$0.7$\pm$1.4)\%\\
50--90&0.92$\pm$0.04$\pm$0.06 & 0.77$\pm$0.03$\pm$0.05 & 0.82$\pm$0.03$\pm$0.05 & 0.85$\pm$0.01$\pm$0.01 & 0.88$\pm$0.01$\pm$0.01 & 0.87$\pm$0.01$\pm$0.01\\
 & (21.0$\pm$1.0$\pm$1.3)\% & (18.5$\pm$0.7$\pm$1.2)\% & (19.5$\pm$0.6$\pm$1.2)\% & (72.0$\pm$1.1$\pm$1.1)\% & (74.0$\pm$0.8$\pm$1.1)\% & (73.2$\pm$0.7$\pm$1.0)\%\\
90--200&0.76$\pm$0.06$\pm$0.05 & 0.90$\pm$0.05$\pm$0.05 & 0.84$\pm$0.04$\pm$0.04 & 0.97$\pm$0.02$\pm$0.02 & 0.90$\pm$0.02$\pm$0.02 & 0.92$\pm$0.01$\pm$0.02\\
 & (17.2$\pm$1.4$\pm$1.0)\% & (20.6$\pm$1.1$\pm$1.2)\% & (19.2$\pm$0.9$\pm$1.0)\% & (73.4$\pm$1.6$\pm$1.3)\% & (69.9$\pm$1.2$\pm$1.3)\% & (71.3$\pm$1.0$\pm$1.2)\%\\
\end{scotch}
}
\end{table*}

\section{Unfolding}\label{sec:unfolding}

The unfolding procedure corrects the measured cross section ratios for effects related to the detector response and the event reconstruction procedures, which can lead to migrations between bins and therefore alter the true distributions. The bin-by-bin migrations are corrected by the response matrices, which quantify the migration probability between the measured and true values of a given observable (jet or \PZ \pt). These matrices are derived in simulation by comparing the final-state objects (jets and leptons) at the pre-reconstruction (``MC-particle") and reconstruction levels.

At the MC-particle level (denoted as ``particle level"), leptons are stable particles from \PZ boson decays, dressed by adding the momenta of all photons within $\Delta R < 0.1$ around the lepton directions. The particle-level jets are formed from stable particles (c$\tau>1\unit{cm}$), except neutrinos, and overlapping leptons from \PZ boson decays, using the same anti-\kt jet algorithm used for reconstructed jets.     

The \PZ boson mass and \pt at the particle level are calculated using the two leptons originating from this boson. The fiducial volume is defined by the particle-level leptons and jets with the same kinematic requirements (\pt, $\eta$, and dilepton invariant mass) used in the measurement.

The response matrix is constructed using MC \zj samples. The reconstructed jets and a pair of electrons or muons are spatially matched to the corresponding particle-level objects by requiring that they are within $\Delta R < 0.2$. In addition, the flavor of the reconstructed jets and the matched particle-level jets must be the same. Events that have reconstructed objects without matched particle-level objects are included in the background category and are subtracted from the sample. The acceptance and efficiency corrections account for other events that have particle-level objects in the fiducial volume but no matching reconstructed objects.

For the inclusive measurement, the acceptance corrections are derived from simulation and defined as the ratios between the number of selected events at the reconstruction level and the number of generated events within the fiducial volume. These acceptance correction factors, which depend on the jet flavor, are applied to the measured cross section ratios.

To unfold the differential distributions, the \tuf package~\cite{Schmitt}, which is based on a least-square fit, is used. The unfolding procedure, which solves for a well-conditioned unfolding problem in this case, is performed without regularization to avoid potential biases toward MC spectra. The data distributions of \zcs, \zbs, and \zj are unfolded simultaneously to include the correlations between the denominator and numerator when deriving the unfolded ratios. The numbers of bins in the unfolded distributions are about half of those used in data to maintain the stability of the unfolding procedure. The combined response matrix used in the simultaneous unfolding is constructed from individual jet category matrices. The \tuf package provides unfolded distributions together with a covariance matrix, which is used to estimate the uncertainties in the unfolded cross section ratios. The unfolding procedure is checked with closure tests and bias studies using MC samples. In the closure test, response matrices are derived using one-half of the sample and the unfolding is performed on the other half. Within statistical uncertainties, the unfolded and MC truth distributions agree with each other. In the bias studies, a pull distribution is constructed by performing the unfolding on a set of ${\approx}100$ MC sub-samples. The unfolding procedure converges and shows no bias.

\section{Systematic uncertainties}\label{sec:syst}

The systematic uncertainties include experimental sources that affect the shape or normalization of templates in the scale factor fits, and the heavy-flavor tagging efficiencies. The unfolded results contain additional uncertainties related to the unfolding procedure. The following systematic uncertainties are considered in the analysis:

{\textit{Jet energy scale and resolution correction:}} The reconstructed jet energy is corrected using a factorized model to compensate for the nonlinear and nonuniform response in the calorimeters as detailed in Ref.~\cite{Khachatryan:2016kdb}. Since the JER is different in data and simulation, the jet energy in simulation is spread to match the resolution observed in data. Both the JES and JER corrections affect the shape of \msv distributions used in the scale factor fits. Therefore, they contribute to the uncertainties in the \zqc and \zbs event yields.

{\textit{Pileup weighting:}} The distribution of the number of pileup events in simulation is weighted to match that in data. The corresponding uncertainty is estimated by varying the total $\Pp\Pp$ inelastic cross section by 4.6\% based on the measurement described in Ref.~\cite{ppinelastic}. Since the shapes of \msv templates are affected by the pileup weighting, this uncertainty source contributes to the \zcs and \zbs event yields as well.

{\textit{Gluon splitting:}} Particles from a pair of collimated \cPqc or \cPqb quarks may end up in the same reconstructed jet, which can affect the shape of \msv template. To quantify the corresponding uncertainties in the scale factor fit, the fraction of MC events with gluon splitting is varied by 50\%, which is about three times the experimental uncertainty in the gluon splitting rate measured at LEP~\cite{gcc,gbb}. The resulting variations in MC \msv shape is propagated to the scale factor fit.

{\textit{Background rates:}} The \ttbar, single top quark, and diboson backgrounds are estimated in simulation using NNLO and NLO cross sections to normalize the event rates. The uncertainties in the \ttbar and diboson background contributions are obtained by varying their production cross sections by 5.5 and 3.2\%, respectively. The uncertainty in the single top backgrounds is ignored because these backgrounds represent a very small fraction (${<}1\%$) of the total event sample.

{\textit{Statistical uncertainties of \msv templates:}} A systematic uncertainty is associated to the limited number of events in the MC samples used to define the template shapes. To estimate the corresponding uncertainty, an ensemble of the \msv templates has been created where the bin contents have been modified by additional statistical fluctuations.

{\textit{Correction of the \bj \msv template:}} This systematic uncertainty is related to the ad hoc shape correction function used to derive the \bj \msv template from control samples in data. This correction, parameterized as a second order polynomial, accounts for the difference in shape of \msv distributions in \ttbar and \zj events. The uncertainty of the shape correction is estimated by changing the polynomial functional forms.

{\textit{Heavy-flavor tagging efficiency:}} The HF tagging efficiencies for \cPqc and \bjs are estimated in simulation and corrected by the efficiency scale factors as described in Section~\ref{subsec:eff}. The systematic uncertainties of the efficiency scale factors of \cjs and \bjs with $30 < \pt < 100\GeV$ and $\abs{\eta} < 2.4$ are ${\approx}3.5\%$ and ${\approx}1.4\%$, respectively~\cite{csv}.

{\textit{Missing transverse momentum selection efficiency:}} This uncertainty source accounts for possible differences in the \ptmiss selection ($\ptmiss < 40\GeV$) efficiencies for \zj and \zhfs events. The effect comes from contributions of semileptonic decays of HF hadrons in \zhfs events, which results in large \ptmiss values. Therefore, the efficiencies tend to be lower for \zhfs events by ${\approx}1\%$ at high jet and \PZ boson \pt regions compared to those of \zj events. An uncertainty of 1.5\% is included in the R(c/j) and R(b/j) differential results for jet (\PZ boson) \pt bins where $\pt>60$ (90)\GeV.

{\textit{PDF and \muR, \muF scale uncertainties:}} These uncertainty sources affect the unfolding correction described in Section~\ref{sec:unfolding}, which is based on the \zj MC samples. The unfolding is performed with different PDF replicas and alternative choices of the renormalization and factorization scales. The uncertainties are obtained from variations of the unfolded results and they are less than 2.5\%, 2.8\%, and 2.9\% in all jet and \PZ \pt bins for R(c/j), R(b/j), and R(c/b), respectively.

{\textit{Parton shower and hadronization model:}} The unfolding procedure is based on response matrices constructed from the \zj simulation sample described in Section~\ref{sec:sample}. This sample uses \PYTHIA to simulate the parton shower and hadronization. An alternative model is provided by the \HERWIGpp generator~\cite{herwig}. The uncertainties in parton shower and hadronization modeling are estimated by comparing the unfolded results using response matrices from those two models. They are less than 3\% for all differential cross section ratios.

Table~\ref{tab:unc_sfs} summarizes the effects of systematic uncertainty sources on the \SFc and \SFb shown in Tables~\ref{tab:sf_fit_1} and~\ref{tab:sf_fit_2}. They are quantified as the differences in quadrature between scale factor uncertainties obtained in two fits: the nominal one where all parameters are allowed to float, and an alternative fit where the nuisance parameter corresponding to the uncertainty source of interest is fixed. The uncertainties from the scale factors and HF tagging efficiency together with the statistical uncertainties of the cross section ratios are listed in Table~\ref{tab:unc_ratios}.

In the unfolded differential results, the uncertainties of the measurements described here are included in the data covariance matrix, which is used to build a least squares fit of the unfolding. An error covariance matrix for the unfolded distributions is estimated. This includes the uncertainties from the data, response matrix, and the unfolding procedure.

\begin{table}[htb]
\topcaption{Systematic uncertainties in the scale factor measurements. The uncertainty ranges correspond to variations across jet and \PZ \pt bins.}
\label{tab:unc_sfs}
\centering
\begin{scotch}{ccc}
 & \SFc & \SFb \\
 \hline
JES, JER               & 1.7--7.4\%         & 0.3--2.1\%\\
Template statistics    & 2.4--6.1\%         & 0.6--2.7\%\\
Gluon splitting        & 2.2--3.9\%         & 0.5--2.0\%\\
Pileup weighting     & 1.6--2.8\%         & 0.3--1.0\%\\
Background uncertainty & 0.3--1.0\%         & 0.4--1.2\%\\
\bj \msv correction   & 0.2--1.6\%         & 0.2--0.8\%\\
Total systematic uncertainty & 4.8--12.5\% & 1.1--4.9\%\\
\end{scotch}
\end{table}

\begin{table}[htb]
\topcaption{The systematic uncertainties in the cross section ratio measurements. The uncertainty ranges correspond to variations across jet and \PZ \pt bins.}
\label{tab:unc_ratios}
\centering
\begin{scotch}{cccc}
 & R(c/j) & R(b/j) & R(c/b)\\
\hline
Scale factor measurement & 5.4--13.8\% & 1.4--4.4\% & 5.6--12.6\%\\
HF tagging               & 3.8--4.6\%  & 1.1--1.5\% & 4.9--6.1\%\\
Statistical uncertainty  & 1.6--7.5\%  & 0.6--3.0\% & 1.8--8.6\%\\
\end{scotch}
\end{table}

\section{Results}\label{sec:result}

The observed and corrected (for the acceptance and efficiency) cross section ratios for the inclusive measurements are summarized in Tables~\ref{tab:rat_results_1} and~\ref{tab:rat_unfolded}, respectively. The measured differential cross section ratios are presented in Appendix~\ref{sec:diff_xsec}, Tables~\ref{tab:rat_results_2} and~\ref{tab:rat_results_3}.

\begin{table*}[ht]
\topcaption{Cross section ratios measured in the electron and muon channels, along with the combined results. The first and second uncertainty values correspond to the statistical and systematic contributions, respectively.}
\label{tab:rat_results_1}
\centering
\cmsTable{
\begin{scotch}{cccc}
  &                                                Electron &          Muon &              Combined\\
\hline
R(c/j) & 0.098 $\pm$ 0.002 $\pm$ 0.009 & 0.094 $\pm$ 0.002 $\pm$ 0.008  & 0.095 $\pm$ 0.002 $\pm$ 0.008\\
R(b/j) & 0.0546 $\pm$ 0.0005 $\pm$ 0.0010 & 0.0538 $\pm$ 0.0004 $\pm$ 0.0010 & 0.0541 $\pm$ 0.0003 $\pm$ 0.0011\\
R(c/b) & 1.80 $\pm$ 0.05 $\pm$ 0.17 & 1.75 $\pm$ 0.04 $\pm$ 0.16 & 1.76 $\pm$ 0.03 $\pm$ 0.16 \\
\end{scotch}
}
\end{table*}

\begin{table*}[ht]
\topcaption{Unfolded cross section ratios in the electron and muon channels, along with the combined results. The first and second uncertainty values correspond to the statistical and systematic contributions, respectively.}
\label{tab:rat_unfolded}
\centering
\begin{scotch}{cccc}
   &                                                Electron &          Muon &              Combined\\
\hline
R(c/j) & 0.105 $\pm$ 0.003 $\pm$ 0.009 & 0.101 $\pm$ 0.002 $\pm$ 0.009 & 0.102 $\pm$ 0.002 $\pm$ 0.009 \\
R(b/j) & 0.0639 $\pm$ 0.0006 $\pm$ 0.0015  & 0.0629 $\pm$ 0.0005 $\pm$ 0.0014 & 0.0633 $\pm$ 0.0004 $\pm$ 0.0015 \\
R(c/b) & 1.65 $\pm$ 0.04 $\pm$ 0.15 & 1.61 $\pm$ 0.04 $\pm$ 0.15 & 1.62 $\pm$ 0.03 $\pm$ 0.15 \\
\end{scotch}
\end{table*}

\begin{figure*}[htb]
\includegraphics[width=0.45\textwidth]{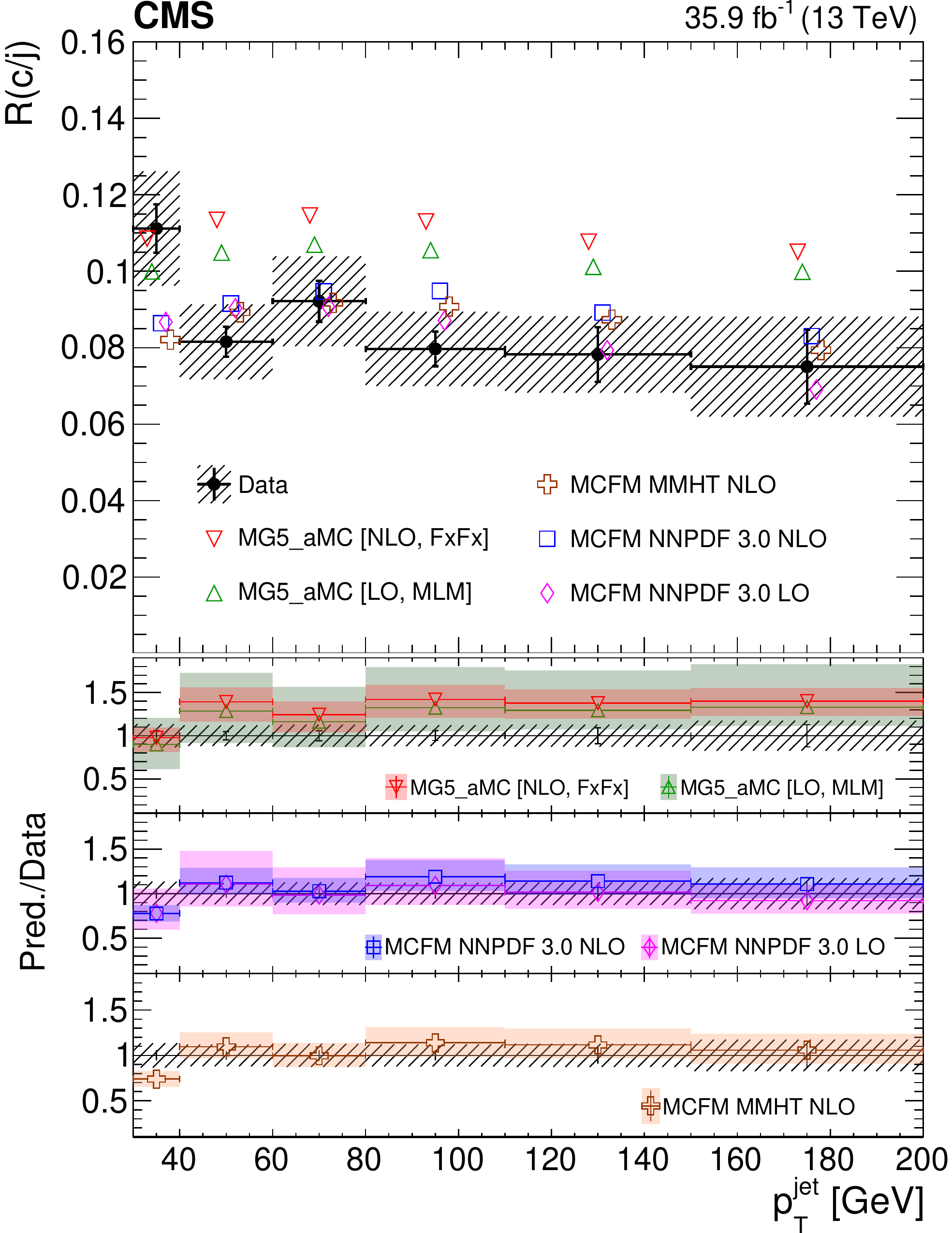}
\includegraphics[width=0.45\textwidth]{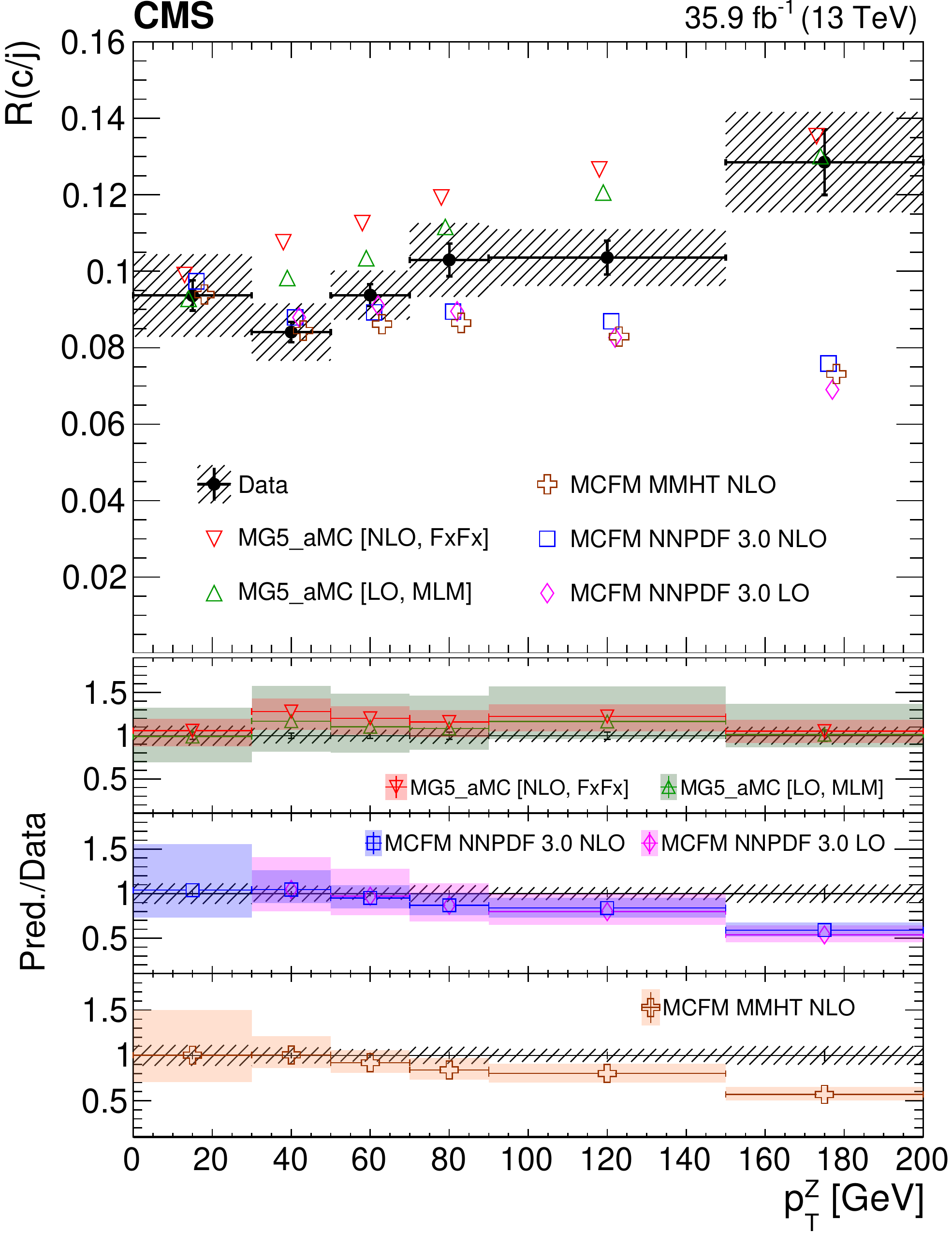}
\caption{Unfolded, particle-level \mgfive, and parton-level \MCFM R(c/j) cross section ratios versus jet (left) and \PZ boson (right) transverse momentum. The vertical error bars for the data points are statistical while the hatched band represents the total uncertainties. The predictions are slightly shifted along the $x$-axis for readability in the upper plots, and their total PDF and scale uncertainties are shown as error bands in the ratio plots.}
\label{fig:diff_cj_comp}
\end{figure*}

\begin{figure*}[htb]
\includegraphics[width=0.45\textwidth]{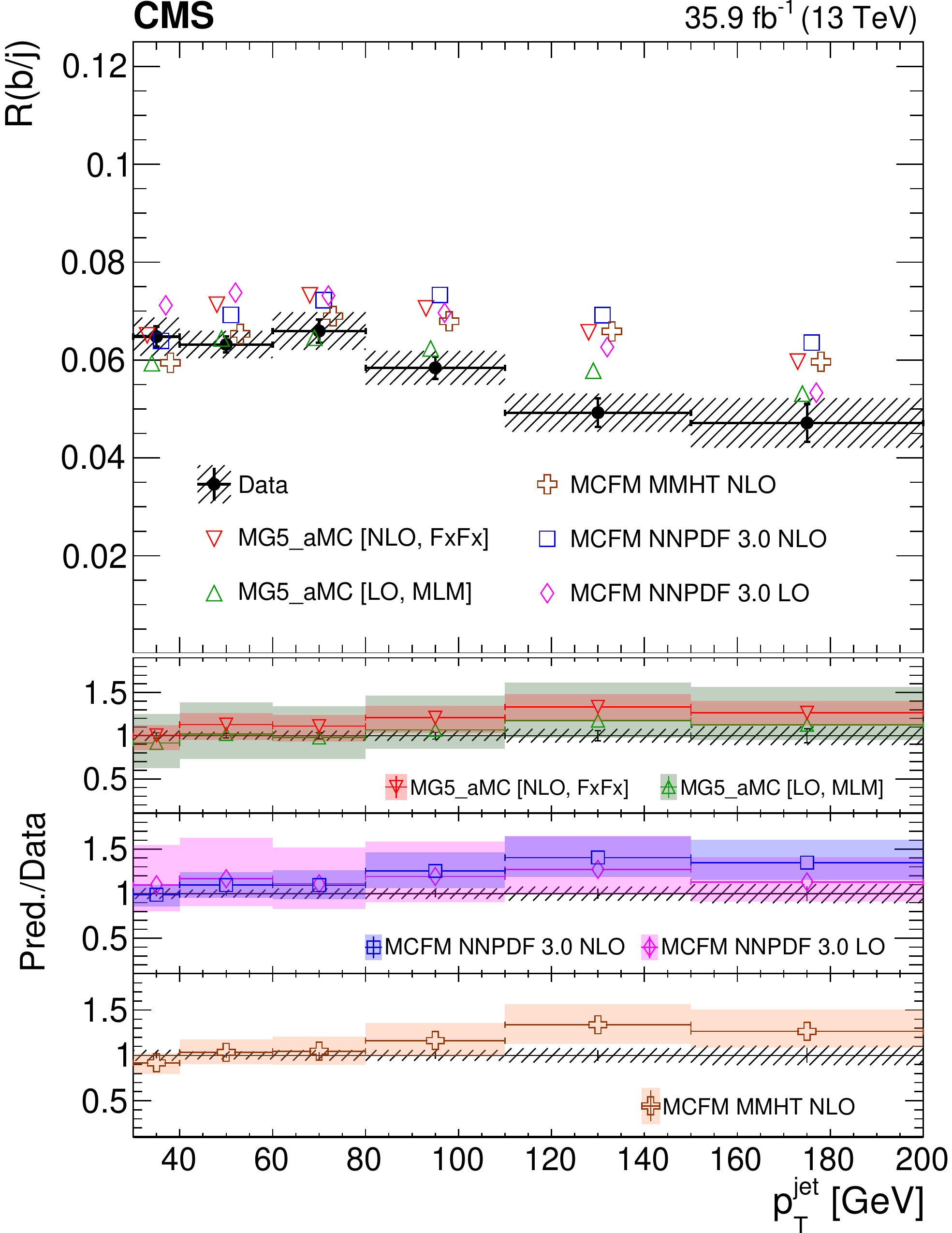}
\includegraphics[width=0.45\textwidth]{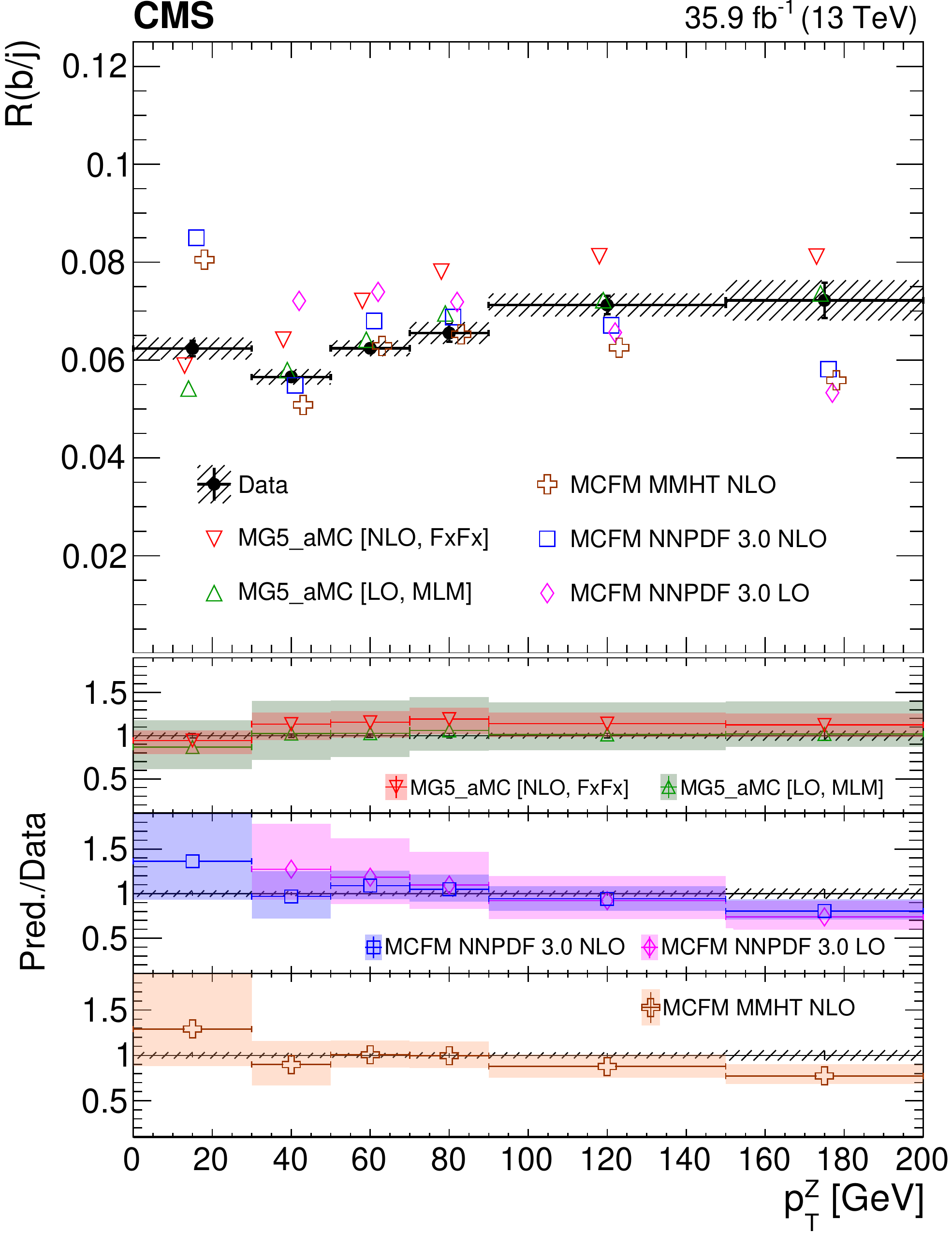}
\caption{Unfolded, particle-level \mgfive, and parton-level \MCFM R(b/j) cross section ratio versus jet (left) and \PZ boson (right) transverse momentum. The vertical error bars for the data points are statistical while the hatched band presents the total uncertainties. The predictions are slightly shifted along the $x$-axis for readability in the upper plots, and their total PDF and scale uncertainties are shown as error bands in the ratio plots.}
\label{fig:diff_bj_comp}
\end{figure*}

\begin{figure*}[htb]
\includegraphics[width=0.45\textwidth]{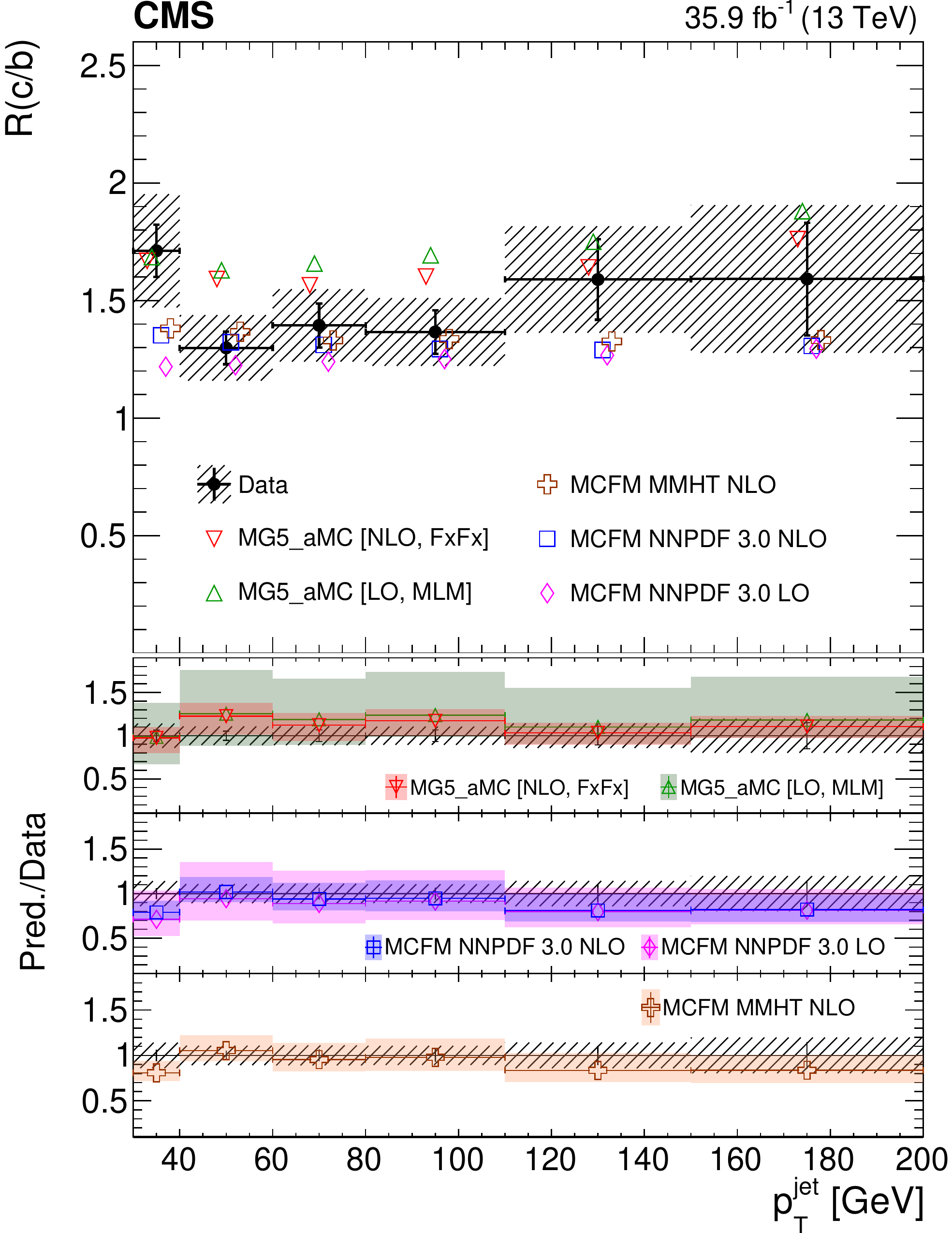}
\includegraphics[width=0.45\textwidth]{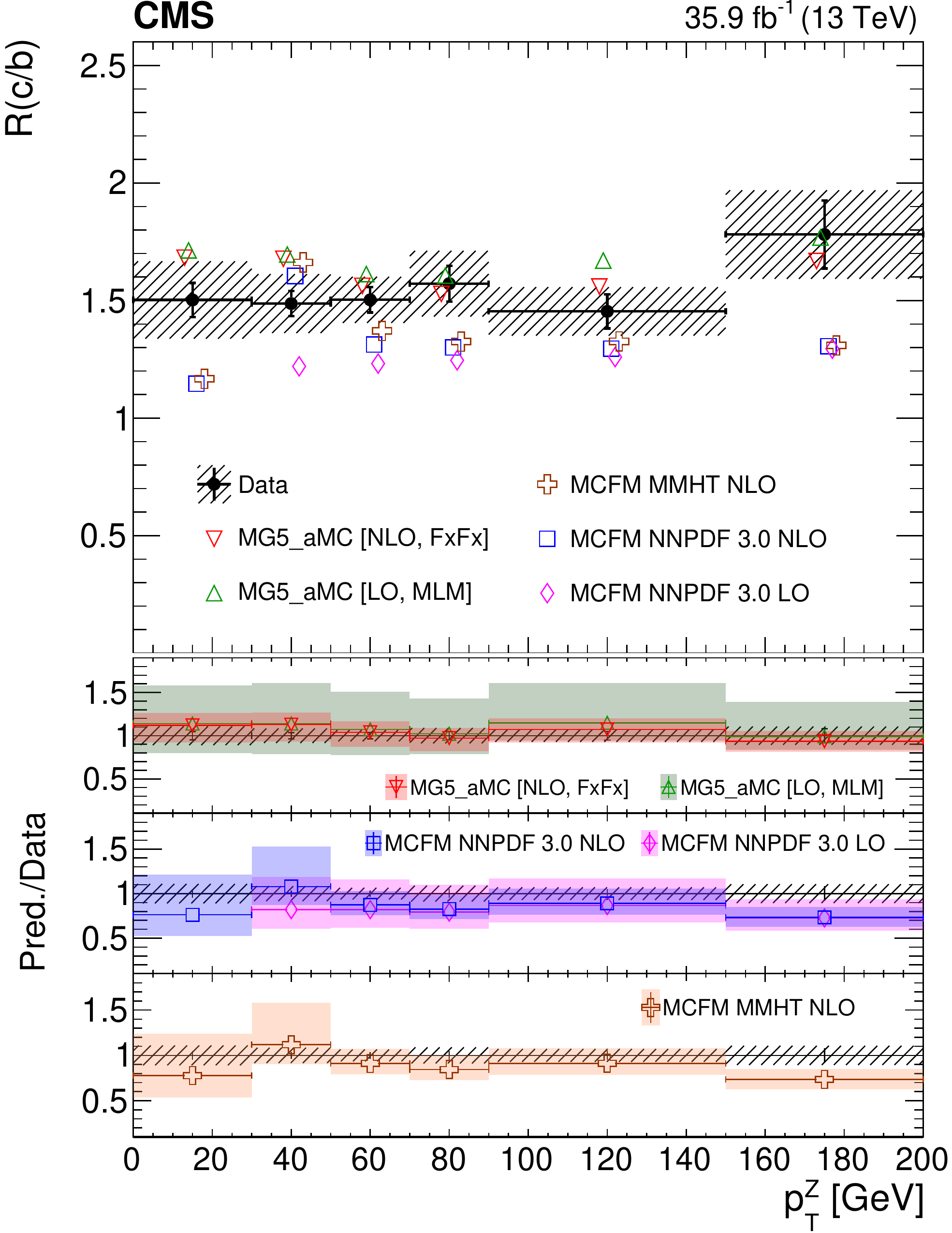}
\caption{Unfolded, particle-level \mgfive, and parton-level \MCFM R(c/b) cross section ratio versus jet (left) and \PZ boson (right) transverse momentum. The vertical error bars for the data points are statistical while the hatched band represents the total uncertainties. The predictions are slightly shifted along the $x$-axis for readability in the upper plots, and their total PDF and scale uncertainties are shown as error bands in the ratio plots.}
\label{fig:diff_cb_comp}
\end{figure*}

The unfolded differential cross section ratios, R(c/j), R(b/j), and R(c/b), versus the jet and \PZ boson \pt are shown in Figs.~\ref{fig:diff_cj_comp},~\ref{fig:diff_bj_comp}, and ~\ref{fig:diff_cb_comp}, respectively. The results are compared with predictions from the \mgfive and \MCFM programs~\cite{MCFM1,MCFM2,MCFM3}, both at LO and NLO. The renormalization and factorization scales in the matrix element and the PDF uncertainties are included in these predictions. For the former, the scales are varied between 0.5 and 2 times their nominal value such that the \muR/\muF ratio is kept between 0.5 and 2. This conventional choice is implemented in the CMS-default settings for generating samples to estimate the theoretical \mgfive LO and NLO cross sections. The uncertainty due to the scales is taken to be the envelope of these predictions. In addition, for the \MCFM calculation, the constraint on the \muR/\muF ratio is dropped, \ie, the scales are varied independently. This more conservative choice is motivated by the fact that the \zhfs cross sections as functions of the renormalization and factorization scales have opposite trends~\cite{Campbell:2003dd}. The \MCFM error bands in Figs.~\ref{fig:diff_cj_comp}--\ref{fig:diff_cb_comp} correspond to this choice of scale variation. The uncertainties due to the scales in the cross section ratios are obtained by adding uncertainties in the numerator and denominator in quadrature, \ie, they are assumed to be uncorrelated. The PDF uncertainty is evaluated by changing the replicas of the PDF set.

The LO cross sections are computed using \mgfive interfaced with \PYTHIA through the \kt-MLM matching scheme~\cite{Alwall:2007fs,Alwall:2008qv}. The LO matrix element calculations include processes with up to 4 outgoing partons. The NNPDF 3.0 LO PDF set is used and the matching scale together with the strong coupling constant \alpS at the \PZ boson mass are set at 19\GeV and 0.130, respectively. The multileg \mgfive generator interfaced with \PYTHIA using the FxFx matching scheme evaluates the cross section ratios at NLO precision. The choice of parameters is described in Section~\ref{sec:sample}.

The \MCFM generator is used to perform calculations of the cross sections and cross section ratios at the parton level in the 5FS. The \zj cross sections are evaluated by a simple cone algorithm with a radius of 0.4 (\ie, partons are merged if the distances, $\Delta R$, between them are less than 0.4). The central values for the cross sections are evaluated at \muR and \muF set to the mass of the \PZ boson. In addition, the NLO \MCFM results are shown for two PDFs, NNPDF 3.0 and MMHT14~\cite{MMHT}, along with the \MCFM LO cross section ratios. The values of \alpS are taken from those PDFs. Table~\ref{tab:rat_prediction} shows the predicted inclusive cross section ratios from \mgfive and \MCFM.

\begin{table*}[htb]
\topcaption{Predicted cross section ratios from \mgfive and \MCFM at LO and NLO accuracy. The first and second sets of uncertainties correspond to PDF and scale variations, respectively. The scale uncertainties for \MCFM with \muR/\muF ratio kept between 0.5 and 2 are in the parentheses.}
\label{tab:rat_prediction}
\centering
\cmsTable{
\begin{scotch}{ccccc}
& \mgfive MG5\_aMC (NLO, FxFx) & \MCFM (NLO) & \mgfive (LO, MLM) & \MCFM (LO) \\
\hline
\rule{0pt}{3ex}
R(c/j) & 0.111 $\pm$ 0.003 $^{+0.010}_{-0.011}$ & 0.090 $\pm$ 0.003 $^{+0.010}_{-0.012}$ ($^{+0.008}_{-0.007}$) & 0.103 $\pm$ 0.003 $^{+0.028}_{-0.026}$ & 0.087 $\pm$ 0.003 $^{+0.025}_{-0.022}$\\
\rule{0pt}{3ex}
R(b/j) & 0.067 $\pm$ 0.002 $\pm$ 0.006 & 0.068 $\pm$ 0.002 $^{+0.008}_{-0.011}$ ($\pm$0.006) & 0.062 $\pm$ 0.002 $^{+0.018}_{-0.015}$ & 0.071 $\pm$ 0.002 $^{+0.023}_{-0.021}$ \\
\rule{0pt}{3ex}
R(c/b) & 1.64 $\pm$ 0.05 $^{+0.15}_{-0.16}$ & 1.33 $\pm$ 0.04 $^{+0.16}_{-0.21}$ ($^{+0.10}_{-0.12}$) & 1.67 $\pm$ 0.06 $^{+0.54}_{-0.40}$ & 1.20 $\pm$ 0.04 $^{+0.42}_{-0.38}$ \\
\end{scotch}
}
\end{table*}

A few comments are in order when comparing data with various predictions. The \mgfive predictions for the cross section ratios are higher in most of the bins, although still compatible with the data given the large uncertainties, except for the R(c/j) versus jet \pt, where the deviations are more pronounced. The data are better described with \mgfive at LO compared to \mgfive at NLO. These observations are similar to those reported in previous measurements at 8\TeV~\cite{ZbCMS,ZCciemat}. The \MCFM predictions for R(c/j) and R(b/j) disagree with data at high jet and \PZ \pt, except for R(c/j) versus jet \pt, where in general there is good agreement with LO or NLO calculations, and for both PDFs considered. For R(c/b), however, all theoretical predictions are consistent with the measured ratios, except for the \MCFM prediction for the highest \PZ boson \pt bin.  The difference between the parton- and particle-level jets may affect the \MCFM predictions, although the corresponding effects are significantly reduced or vanish in the cross section ratios. Alternatively, higher order pQCD calculations might be needed to describe the data.

\clearpage
\section{Summary}{\label{sec:sum}}
Ratios of cross sections, $\sigma(\zcs)/\sigma(\zj)$, $\sigma(\zbs)/\sigma(\zj)$, and $\sigma(\zcs)/\sigma(\zbs)$ in the associated production of a \PZ boson with at least one charm or bottom quark jet have been measured in proton-proton collisions at $\sqrt s=13\TeV$ using 35.9\fbinv of data collected by the CMS experiment at the LHC. The fiducial volume of the measurement is defined by $\pt>30\GeV$ and $|\eta|<2.4$ for the jets, where \pt and $\eta$ represent transverse momentum and pseudorapidity, respectively. The \PZ bosons are selected within the mass range of 71 and 111\GeV requiring leptons (electrons or muons) with $\pt > 25\GeV$ and $\abs{\eta} < 2.4$. The measured values are $\sigma(\zcs)/\sigma(\zj) = 0.102 \pm 0.002 \pm 0.009$, $\sigma(\zbs)/\sigma(\zj) = 0.0633 \pm 0.0004 \pm 0.0015$, and $\sigma(\zcs)/\sigma(\zbs) = 1.62 \pm 0.03 \pm 0.15$. Results for the inclusive and differential cross section ratios as functions of jet and \PZ boson transverse momentum are compared with predictions from leading and next-to-leading order perturbative quantum chromodynamics calculations. These are the first results of this kind at 13\TeV.

\begin{acknowledgments}
We congratulate our colleagues in the CERN accelerator departments for the excellent performance of the LHC and thank the technical and administrative staffs at CERN and at other CMS institutes for their contributions to the success of the CMS effort. In addition, we gratefully acknowledge the computing centers and personnel of the Worldwide LHC Computing Grid for delivering so effectively the computing infrastructure essential to our analyses. Finally, we acknowledge the enduring support for the construction and operation of the LHC and the CMS detector provided by the following funding agencies: BMBWF and FWF (Austria); FNRS and FWO (Belgium); CNPq, CAPES, FAPERJ, FAPERGS, and FAPESP (Brazil); MES (Bulgaria); CERN; CAS, MoST, and NSFC (China); COLCIENCIAS (Colombia); MSES and CSF (Croatia); RPF (Cyprus); SENESCYT (Ecuador); MoER, ERC IUT, PUT and ERDF (Estonia); Academy of Finland, MEC, and HIP (Finland); CEA and CNRS/IN2P3 (France); BMBF, DFG, and HGF (Germany); GSRT (Greece); NKFIA (Hungary); DAE and DST (India); IPM (Iran); SFI (Ireland); INFN (Italy); MSIP and NRF (Republic of Korea); MES (Latvia); LAS (Lithuania); MOE and UM (Malaysia); BUAP, CINVESTAV, CONACYT, LNS, SEP, and UASLP-FAI (Mexico); MOS (Montenegro); MBIE (New Zealand); PAEC (Pakistan); MSHE and NSC (Poland); FCT (Portugal); JINR (Dubna); MON, RosAtom, RAS, RFBR, and NRC KI (Russia); MESTD (Serbia); SEIDI, CPAN, PCTI, and FEDER (Spain); MOSTR (Sri Lanka); Swiss Funding Agencies (Switzerland); MST (Taipei); ThEPCenter, IPST, STAR, and NSTDA (Thailand); TUBITAK and TAEK (Turkey); NASU (Ukraine); STFC (United Kingdom); DOE and NSF (USA).

\hyphenation{Rachada-pisek} Individuals have received support from the Marie-Curie program and the European Research Council and Horizon 2020 Grant, contract Nos.\ 675440, 752730, and 765710 (European Union); the Leventis Foundation; the A.P.\ Sloan Foundation; the Alexander von Humboldt Foundation; the Belgian Federal Science Policy Office; the Fonds pour la Formation \`a la Recherche dans l'Industrie et dans l'Agriculture (FRIA-Belgium); the Agentschap voor Innovatie door Wetenschap en Technologie (IWT-Belgium); the F.R.S.-FNRS and FWO (Belgium) under the ``Excellence of Science -- EOS" -- be.h project n.\ 30820817; the Beijing Municipal Science \& Technology Commission, No. Z191100007219010; the Ministry of Education, Youth and Sports (MEYS) of the Czech Republic; the Deutsche Forschungsgemeinschaft (DFG) under Germany’s Excellence Strategy -- EXC 2121 ``Quantum Universe" -- 390833306; the Lend\"ulet (``Momentum") Program and the J\'anos Bolyai Research Scholarship of the Hungarian Academy of Sciences, the New National Excellence Program \'UNKP, the NKFIA research grants 123842, 123959, 124845, 124850, 125105, 128713, 128786, and 129058 (Hungary); the Council of Science and Industrial Research, India; the HOMING PLUS program of the Foundation for Polish Science, cofinanced from European Union, Regional Development Fund, the Mobility Plus program of the Ministry of Science and Higher Education, the National Science Center (Poland), contracts Harmonia 2014/14/M/ST2/00428, Opus 2014/13/B/ST2/02543, 2014/15/B/ST2/03998, and 2015/19/B/ST2/02861, Sonata-bis 2012/07/E/ST2/01406; the National Priorities Research Program by Qatar National Research Fund; the Ministry of Science and Education, grant no. 14.W03.31.0026 (Russia); the Programa Estatal de Fomento de la Investigaci{\'o}n Cient{\'i}fica y T{\'e}cnica de Excelencia Mar\'{\i}a de Maeztu, grant MDM-2015-0509 and the Programa Severo Ochoa del Principado de Asturias; the Thalis and Aristeia programs cofinanced by EU-ESF and the Greek NSRF; the Rachadapisek Sompot Fund for Postdoctoral Fellowship, Chulalongkorn University and the Chulalongkorn Academic into Its 2nd Century Project Advancement Project (Thailand); the Kavli Foundation; the Nvidia Corporation; the SuperMicro Corporation; the Welch Foundation, contract C-1845; and the Weston Havens Foundation (USA). \end{acknowledgments}

\bibliography{auto_generated}

\clearpage
\appendix

\section{Post-fit \texorpdfstring{\msv}{M[SV]} distributions in the exclusive jet \texorpdfstring{\pt}{pt} bins}\label{sec:msv_dist}
\begin{figure*}[hb]
\centering
\includegraphics[width=0.45\textwidth]{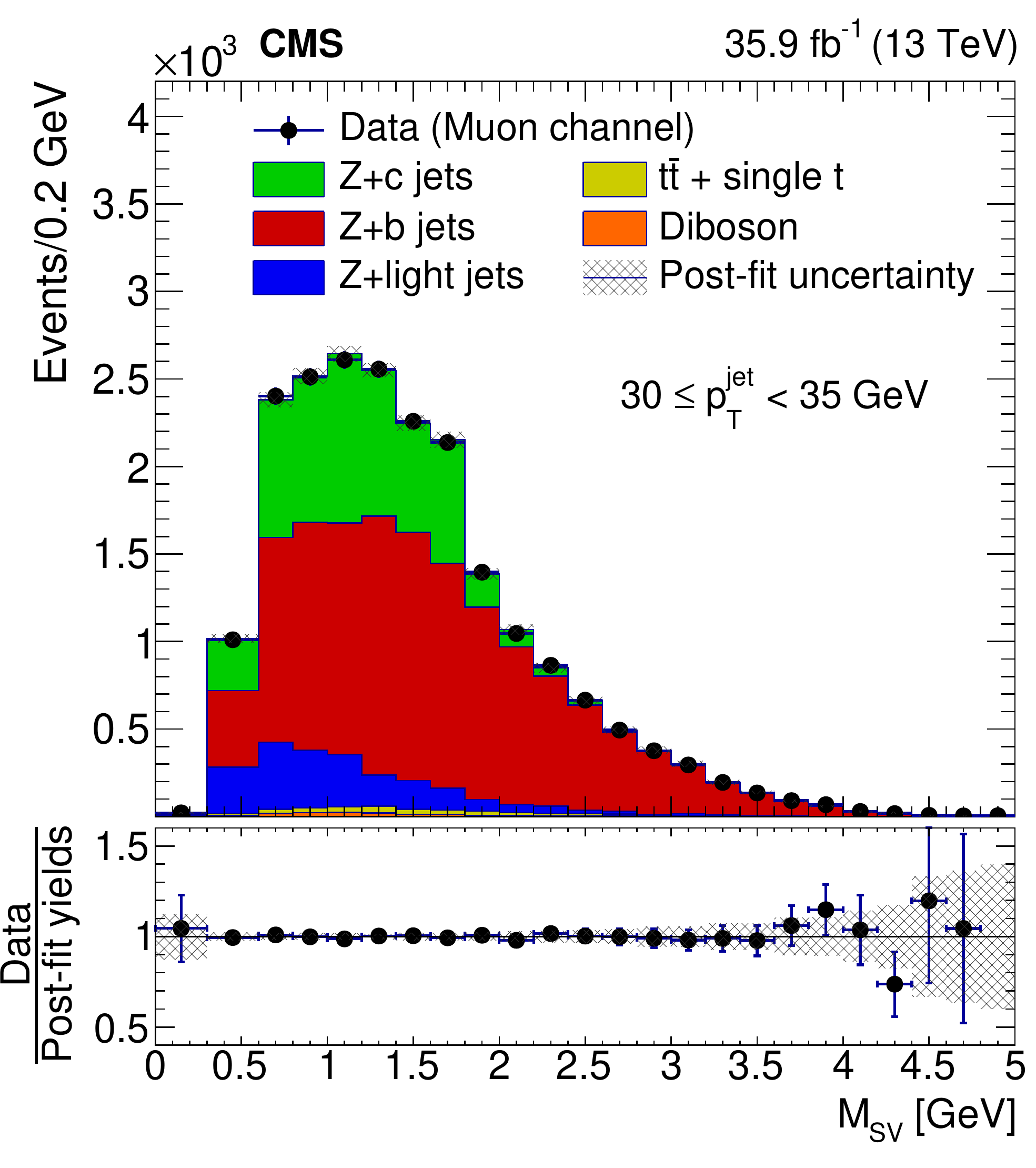}
\includegraphics[width=0.45\textwidth]{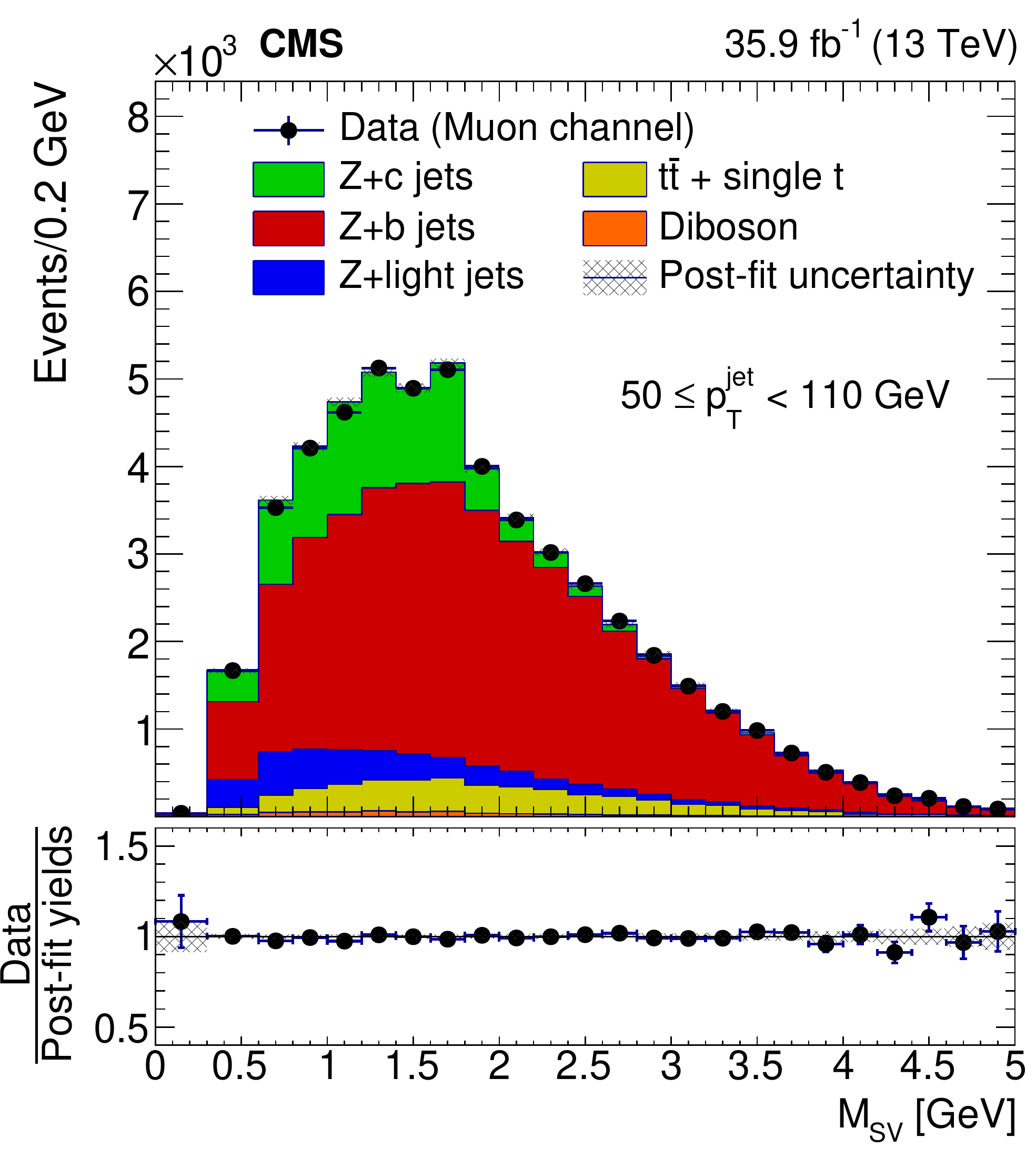}
\includegraphics[width=0.45\textwidth]{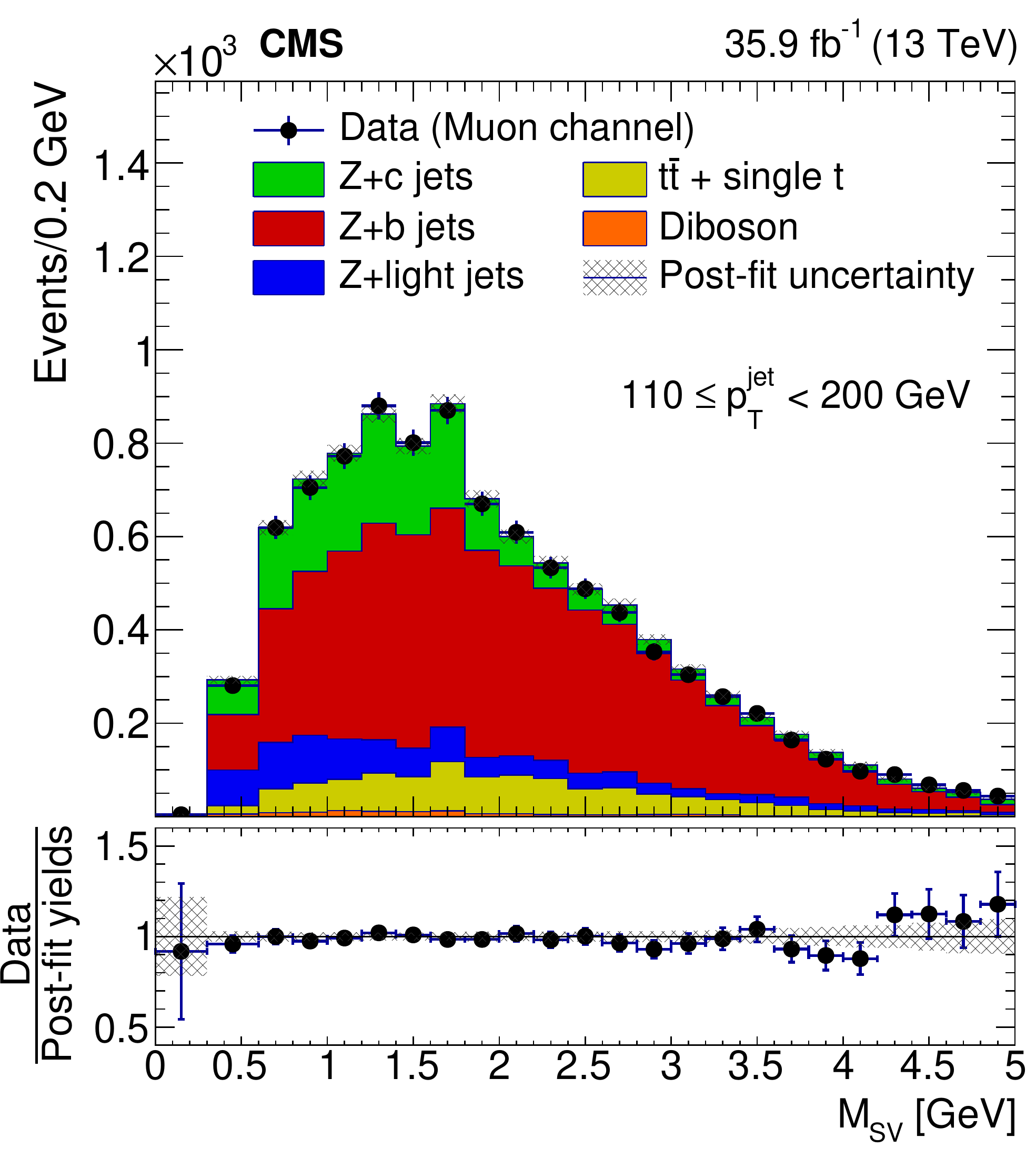}
\caption{Secondary vertex invariant mass distributions for jet \pt bins 30--35\GeV, 50--110\GeV, and 110-200\GeV in the muon channel derived from fits using the corresponding jet \pt binned \zhfs data samples. The post-fit uncertainty bands indicate the total uncertainties, added in quadrature, of the best-fit values of signal and background process rates.}
\label{fig:fig_postfit_2}
\end{figure*}

\section{Measured differential cross section ratios}\label{sec:diff_xsec}
\ifthenelse{\boolean{cms@external}}{\begin{turnpage}}{\begin{landscape}}
\begin{table*}[htb]
\topcaption{The cross section ratios for the electron, muon, and combined channels in jet \pt bins.}
\label{tab:rat_results_2}
\centering
\cmsTableRot{
\begin{tabular}{|c|c|c|c|c|c|c|c|c|c|}
\hline
\multirow{2}{*}{Jet \pt} & \multicolumn{3}{c|}{Electron} & \multicolumn{3}{c|}{Muon} & \multicolumn{3}{c|}{Combined} \\
\cline{2-10}
& R(c/j) & R(b/j) & R(c/b) & R(c/j) & R(b/j) & R(c/b) & R(c/j) & R(b/j) & R(c/b)\\
\hline
30--35&0.105$\pm$0.006$\pm$0.010&0.0487$\pm$0.0011$\pm$0.0020&2.15$\pm$0.13$\pm$0.28&0.103$\pm$0.004$\pm$0.010&0.0468$\pm$0.0009$\pm$0.0019&2.21$\pm$0.10$\pm$0.28&0.104$\pm$0.003$\pm$0.008&0.0475$\pm$0.0007$\pm$0.0016&2.19$\pm$0.09$\pm$0.23\\
35--40&0.091$\pm$0.006$\pm$0.010&0.0568$\pm$0.0014$\pm$0.0023&1.59$\pm$0.12$\pm$0.22&0.093$\pm$0.005$\pm$0.010&0.0545$\pm$0.0011$\pm$0.0021&1.71$\pm$0.10$\pm$0.22&0.092$\pm$0.004$\pm$0.008&0.0556$\pm$0.0009$\pm$0.0019&1.65$\pm$0.08$\pm$0.18\\
40--50&0.079$\pm$0.005$\pm$0.008&0.0554$\pm$0.0010$\pm$0.0016&1.43$\pm$0.10$\pm$0.18&0.080$\pm$0.004$\pm$0.009&0.0546$\pm$0.0008$\pm$0.0016&1.46$\pm$0.08$\pm$0.18&0.079$\pm$0.003$\pm$0.008&0.0549$\pm$0.0007$\pm$0.0014&1.44$\pm$0.06$\pm$0.16\\
50--60&0.102$\pm$0.004$\pm$0.008&0.0596$\pm$0.0009$\pm$0.0017&1.71$\pm$0.08$\pm$0.14&0.084$\pm$0.003$\pm$0.007&0.0611$\pm$0.0008$\pm$0.0022&1.38$\pm$0.06$\pm$0.12&0.092$\pm$0.003$\pm$0.007&0.0606$\pm$0.0007$\pm$0.0020&1.51$\pm$0.05$\pm$0.12\\
60--70&0.093$\pm$0.004$\pm$0.007&0.0591$\pm$0.0009$\pm$0.0019&1.57$\pm$0.07$\pm$0.13&0.084$\pm$0.003$\pm$0.007&0.0600$\pm$0.0007$\pm$0.0024&1.39$\pm$0.06$\pm$0.12&0.088$\pm$0.003$\pm$0.006&0.0597$\pm$0.0006$\pm$0.0022&1.47$\pm$0.05$\pm$0.11\\
70--90&0.091$\pm$0.004$\pm$0.007&0.0580$\pm$0.0009$\pm$0.0018&1.56$\pm$0.08$\pm$0.13&0.078$\pm$0.003$\pm$0.006&0.0592$\pm$0.0008$\pm$0.0023&1.32$\pm$0.06$\pm$0.11&0.083$\pm$0.003$\pm$0.006&0.0587$\pm$0.0007$\pm$0.0021&1.42$\pm$0.05$\pm$0.11\\
90--110&0.093$\pm$0.004$\pm$0.007&0.0541$\pm$0.0009$\pm$0.0017&1.71$\pm$0.08$\pm$0.14&0.075$\pm$0.003$\pm$0.006&0.0550$\pm$0.0007$\pm$0.0021&1.36$\pm$0.06$\pm$0.12&0.082$\pm$0.003$\pm$0.006&0.0546$\pm$0.0007$\pm$0.0020&1.50$\pm$0.06$\pm$0.11\\
110--130&0.070$\pm$0.009$\pm$0.007&0.0529$\pm$0.0022$\pm$0.0025&1.32$\pm$0.18$\pm$0.19&0.084$\pm$0.007$\pm$0.008&0.0487$\pm$0.0017$\pm$0.0025&1.73$\pm$0.17$\pm$0.21&0.078$\pm$0.006$\pm$0.006&0.0499$\pm$0.0014$\pm$0.0024&1.57$\pm$0.13$\pm$0.16\\
130--150&0.073$\pm$0.009$\pm$0.007&0.0510$\pm$0.0022$\pm$0.0024&1.43$\pm$0.20$\pm$0.20&0.081$\pm$0.007$\pm$0.007&0.0458$\pm$0.0017$\pm$0.0023&1.76$\pm$0.18$\pm$0.22&0.078$\pm$0.006$\pm$0.006&0.0474$\pm$0.0014$\pm$0.0023&1.64$\pm$0.14$\pm$0.17\\
150--200&0.064$\pm$0.008$\pm$0.006&0.0447$\pm$0.0019$\pm$0.0021&1.43$\pm$0.20$\pm$0.20&0.081$\pm$0.007$\pm$0.007&0.0454$\pm$0.0017$\pm$0.0023&1.78$\pm$0.18$\pm$0.22&0.074$\pm$0.006$\pm$0.006&0.0448$\pm$0.0014$\pm$0.0021&1.65$\pm$0.14$\pm$0.17\\
\hline
\end{tabular}
}
\end{table*}

\begin{table}
\topcaption{The cross section ratios in the electron, muon and combined channels in the \PZ boson \pt bins.}
\label{tab:rat_results_3}
\centering{
\cmsTableRot{
\begin{tabular}{|c|c|c|c|c|c|c|c|c|c|}
\hline
\multirow{2}{*}{\PZ boson \pt} &\multicolumn{3}{c|}{Electron} & \multicolumn{3}{c|}{Muon} & \multicolumn{3}{c|}{Combined} \\
\cline{2-10}
& R(c/j) & R(b/j) & R(c/b) & R(c/j) & R(b/j) & R(c/b) & R(c/j) & R(b/j) & R(c/b)\\
\hline
0--20&0.066$\pm$0.003$\pm$0.010&0.0371$\pm$0.0008$\pm$0.0011&1.77$\pm$0.10$\pm$0.26&0.061$\pm$0.003$\pm$0.009&0.0348$\pm$0.0006$\pm$0.0010&1.77$\pm$0.09$\pm$0.24&0.064$\pm$0.002$\pm$0.009&0.0357$\pm$0.0005$\pm$0.0010&1.78$\pm$0.07$\pm$0.25\\
20--30&0.095$\pm$0.005$\pm$0.014&0.0505$\pm$0.0010$\pm$0.0015&1.89$\pm$0.11$\pm$0.28&0.088$\pm$0.004$\pm$0.012&0.0509$\pm$0.0008$\pm$0.0015&1.73$\pm$0.08$\pm$0.23&0.091$\pm$0.003$\pm$0.013&0.0509$\pm$0.0007$\pm$0.0014&1.80$\pm$0.07$\pm$0.25\\
30--40&0.089$\pm$0.004$\pm$0.009&0.0502$\pm$0.0009$\pm$0.0012&1.78$\pm$0.09$\pm$0.19&0.083$\pm$0.003$\pm$0.008&0.0500$\pm$0.0007$\pm$0.0012&1.67$\pm$0.07$\pm$0.17&0.085$\pm$0.003$\pm$0.008&0.0501$\pm$0.0006$\pm$0.0012&1.71$\pm$0.06$\pm$0.17\\
40--50&0.086$\pm$0.004$\pm$0.008&0.0537$\pm$0.0010$\pm$0.0013&1.60$\pm$0.09$\pm$0.17&0.084$\pm$0.003$\pm$0.008&0.0547$\pm$0.0008$\pm$0.0013&1.53$\pm$0.07$\pm$0.15&0.084$\pm$0.003$\pm$0.008&0.0543$\pm$0.0007$\pm$0.0013&1.55$\pm$0.06$\pm$0.15\\
50--60&0.102$\pm$0.005$\pm$0.008&0.0593$\pm$0.0010$\pm$0.0012&1.71$\pm$0.09$\pm$0.15&0.090$\pm$0.004$\pm$0.007&0.0598$\pm$0.0008$\pm$0.0012&1.50$\pm$0.07$\pm$0.13&0.095$\pm$0.003$\pm$0.007&0.0596$\pm$0.0007$\pm$0.0011&1.59$\pm$0.06$\pm$0.13\\
60--70&0.103$\pm$0.005$\pm$0.008&0.0631$\pm$0.0011$\pm$0.0013&1.64$\pm$0.08$\pm$0.14&0.091$\pm$0.004$\pm$0.007&0.0648$\pm$0.0009$\pm$0.0013&1.41$\pm$0.06$\pm$0.12&0.096$\pm$0.003$\pm$0.007&0.0641$\pm$0.0007$\pm$0.0012&1.50$\pm$0.05$\pm$0.12\\
70--90&0.112$\pm$0.005$\pm$0.008&0.0627$\pm$0.0011$\pm$0.0012&1.79$\pm$0.09$\pm$0.15&0.098$\pm$0.004$\pm$0.008&0.0658$\pm$0.0009$\pm$0.0013&1.49$\pm$0.07$\pm$0.13&0.104$\pm$0.003$\pm$0.008&0.0646$\pm$0.0008$\pm$0.0012&1.60$\pm$0.06$\pm$0.13\\
90--120&0.096$\pm$0.008$\pm$0.007&0.0724$\pm$0.0017$\pm$0.0019&1.32$\pm$0.12$\pm$0.13&0.115$\pm$0.006$\pm$0.008&0.0690$\pm$0.0014$\pm$0.0019&1.67$\pm$0.10$\pm$0.14&0.107$\pm$0.005$\pm$0.007&0.0704$\pm$0.0012$\pm$0.0018&1.52$\pm$0.08$\pm$0.12\\
120--150&0.099$\pm$0.008$\pm$0.007&0.0755$\pm$0.0018$\pm$0.0020&1.31$\pm$0.12$\pm$0.13&0.116$\pm$0.007$\pm$0.008&0.0685$\pm$0.0015$\pm$0.0019&1.69$\pm$0.11$\pm$0.15&0.109$\pm$0.005$\pm$0.007&0.0712$\pm$0.0013$\pm$0.0018&1.53$\pm$0.08$\pm$0.12\\
150--200&0.114$\pm$0.009$\pm$0.009&0.0710$\pm$0.0017$\pm$0.0019&1.61$\pm$0.14$\pm$0.15&0.136$\pm$0.008$\pm$0.010&0.0708$\pm$0.0015$\pm$0.0019&1.93$\pm$0.12$\pm$0.17&0.127$\pm$0.006$\pm$0.009&0.0709$\pm$0.0013$\pm$0.0018&1.79$\pm$0.10$\pm$0.14\\
\hline
\end{tabular}
}
}
\end{table}
\ifthenelse{\boolean{cms@external}}{\end{turnpage}}{\end{landscape}}

\cleardoublepage \section{The CMS Collaboration \label{app:collab}}\begin{sloppypar}\hyphenpenalty=5000\widowpenalty=500\clubpenalty=5000\vskip\cmsinstskip
\textbf{Yerevan Physics Institute, Yerevan, Armenia}\\*[0pt]
A.M.~Sirunyan$^{\textrm{\dag}}$, A.~Tumasyan
\vskip\cmsinstskip
\textbf{Institut f\"{u}r Hochenergiephysik, Wien, Austria}\\*[0pt]
W.~Adam, F.~Ambrogi, T.~Bergauer, J.~Brandstetter, M.~Dragicevic, J.~Er\"{o}, A.~Escalante~Del~Valle, M.~Flechl, R.~Fr\"{u}hwirth\cmsAuthorMark{1}, M.~Jeitler\cmsAuthorMark{1}, N.~Krammer, I.~Kr\"{a}tschmer, D.~Liko, T.~Madlener, I.~Mikulec, N.~Rad, J.~Schieck\cmsAuthorMark{1}, R.~Sch\"{o}fbeck, M.~Spanring, D.~Spitzbart, W.~Waltenberger, C.-E.~Wulz\cmsAuthorMark{1}, M.~Zarucki
% ...
\vskip\cmsinstskip
\dag: Deceased\\
1:  Also at Vienna University of Technology, Vienna, Austria\\
% ...
\end{sloppypar}
%%% END EDITABLE REGION %%%
\end{document}